\documentclass[%
 reprint,
superscriptaddress,
 amsmath,amssymb,
 aps,pre,
]{revtex4-2}

\usepackage{graphicx}
\usepackage{dcolumn}
\usepackage{bm}
\usepackage{hyperref}
\usepackage{placeins}
\usepackage{xcolor}
\usepackage{hyperref}
\hypersetup{
    hidelinks,
    colorlinks=true,
    citecolor=blue,
    linkcolor=blue,
    urlcolor=blue,
    }
\usepackage{amsmath}
\usepackage{amssymb}
\usepackage{gensymb}
\usepackage{subcaption}
\usepackage[normalem]{ulem} 

\begin{document}

\preprint{APS/123-QED}

\title{Bifurcations and multistability in inducible three-gene toggle switch networks}

\author{Rebecca J. Rousseau}
 \email{rroussea@caltech.edu}
\affiliation{Department of Physics, California Institute of Technology, Pasadena, CA 91125}
\author{Rob Phillips} \email{phillips@pboc.caltech.edu}
\affiliation{Department of Physics, California Institute of Technology, Pasadena, CA 91125}
\affiliation{Division of Biology and Biological Engineering, California Institute of Technology, Pasadena, CA 91125}

\date{\today}

\begin{abstract}
Control of transcription presides over a vast array of biological processes, including those mediated by gene regulatory circuits that exhibit multistability. Within these circuits, two- and three-gene network motifs are particularly critical to the repertoire of metabolic and developmental pathways. Theoretical models of these circuits, however, often vary parameters such as dissociation constants, transcription rates, and degradation rates without specifying precisely how these parameters are controlled biologically.  In this study, we examine the role of effector molecules, which can alter the concentrations of the active transcription factors that control regulation, and are ubiquitous in regulatory processes across many biological settings. We specifically consider allosteric regulation in the context of extending the standard bistable switch to three-gene networks, and explore the rich multistable dynamics exhibited in these architectures as a function of effector concentrations. We then analyze how the dynamics evolve under various interpretations of regulatory circuit mechanics, underlying inducer activity, and perturbations thereof. Notably, the biological mechanism by which we model effector control over dual-function proteins transforms not only the phenotypic trend of dynamic tuning but also the set of available dynamic regimes. In this way, we determine key parameters and regulatory features that drive phenotypic decisions, and offer an experimentally tunable structure for encoding inducible multistable behavior arising from both single and dual-function allosteric transcription factors.
\end{abstract}

\maketitle

\section{Introduction}

Biological processes rely on intricate networks of gene interactions that together encode a multitude of possible cellular functions. Often, these genes produce proteins, called transcription factors, that alter the expression of other genes. Whether considering resource consumption in \textit{E. coli}~\cite{Ozbudak:2004} or stages of animal development such as digit formation~\cite{DigitFormation:2012, DigitFormation:2014, DigitFormationPerspective:2014}, vulval development~\cite{VulvaSternberg:1989, VulvaHoyos:2011}, or stem cell differentiation~\cite{StemCellTakahashi:2006, StemCellHanna:2010, StemCellLevine:2017}, such biological systems can precisely tune relevant gene interactions toward many possible distinct phenotypes. This inherent multistability is crucial to cell function, allowing signaling from the environment, metabolic resource constraints, or other sources to induce systems toward what at times can be dramatically different end states.

Determining from experimental data alone the exact regulatory interactions that collectively drive multistable dynamics, however, is another matter. Indeed, the sensitivities of dynamics and steady state outcomes to both internal and external tuning are challenging to disentangle. To better understand these dynamics, we seek to explore motifs that establish the building blocks for multistability in higher-order networks.

One regulatory motif that has emerged as an integral component of natural and synthetic genetic circuit design is the bistable toggle switch, in which each gene produces a protein that represses expression of the other. Study of the lysis vs. lysogeny decision in bacteriophage lambda over the 1960-80s established the significance of this motif experimentally~\cite{JohnsonPtashne:1981, Ptashne2002, Ptashne2004,GoldingSwitch:2011}, and subsequent research has modeled such motifs extensively using Hill models~\cite{Gardner:2000, CherryAdler:2000, HuangEnver:2007}. As a two-state model, however, the bistable switch sheds only partial light on multistability in phenotypic expression.

Insights into the emergence of multistable dynamics in larger networks rely not only on the choice of model but also on the approach used. Existing literature on multistable cell fates during development, for example, has made significant strides by describing gene network dynamics within a Waddington landscape~\cite{Waddington1957} and discussing the dynamic thresholds, or bifurcations, that can emerge as the underlying topology of an attractor system evolves~\cite{CorsonSiggia:2012, CorsonSiggia:2017, RandCS:2021,Rand:2022,RajuSiggia:2023}. These studies largely use geometric approaches that either coarse-grain representations to a set of network-free flow equations, or focus on systems that obey a gradient-like description. Given that many nontrivial networks contain feedback and cyclic interactions that break the monotonicity required for gradient-like behavior, however, other studies focus on the rotational effects that impact a system's trajectory through the potential landscape~\cite{Wang:2011,Stumpf:2018,Wang:2020}. These perspectives have been instrumental in characterizing a range of regulatory networks, often using abstract topological parametrizations and catastrophe theory. In this study, we build upon these rich traditions by directly linking bifurcation analysis of non-gradient multistable networks to experimentally accessible tuning knobs. In particular, we incorporate a concrete thermodynamic interpretation of induction to examine its explicit role in shaping multistable dynamics.

In fact, the activity of transcription factors often depends on the presence or absence of allosteric effector molecules~\cite{Gerhart1962, Monod1963,Monod1965,o1980equilibrium, DalyMatthews1986exptind,Martins2011,Marzen2013,Changeux2013,Gerhart2014,Phillips:2020}. When targeting a repressor, for example, an effector can act as an inducer by binding and stabilizing a protein in its inactive configuration, which suppresses its function and induces expression. Certain transcription factors are also known to have context-dependent behavior, acting as repressors at one promoter site along the DNA and as activators at another~\cite{Alberts,Perry:1996}. Exploring different interpretations of the mechanistic role of effectors may then reveal critical differences in tuned expression that allow for complementary experimental investigations. 

Accordingly, our aim is to examine the role of the effector concentration as a system parameter, and by tuning this parameter alone, to directly observe corresponding dynamical shifts that occur. This framework offers a perspective with parametrizations that are accessible to the experimentalist while also inherent to cell activity itself. For example, researchers can directly input external non-metabolizable effectors to study the dynamic shifts observed in theory~\cite{Marbach2012}. Cells also naturally leverage one or more effectors to tune dynamics, with two inducers driving cell fate determination in the lac operon and in cell differentiation during development, among other settings~\cite{Santillan2004,HuangEnver:2007}. It is important to recognize, of course, that tuning extracellular effector concentrations does not always exactly correspond to intracellular levels due to internal processes that can increase hypersensitivity, and that the internal concentrations of multiple distinct effectors can at times be codependent~\cite{Kuhlman2007}. Nonetheless, effector concentration remains a valuable measure of a cell's ability to regulate its gene expression dynamics, especially in light of the extensive experimental evidence emphasizing their significance as natural tuning knobs.

To analyze the roles of effector molecules in the emergence of multistability, we specifically study extensions of the toggle switch to three genes. These motifs remain simple enough to visualize and physically interpret, yet also provide an additional layer of tunable gene interaction parameters that allows for more complex multistable dynamics -- in this case, tristability and beyond. In exploring how a set of more than two genes couple to collectively decide among multiple possible cell fates, we also gain insights into how these interactions drive dynamics in larger and more complex networks.

\begin{figure}
    \centering
    \includegraphics[width=0.5\columnwidth]{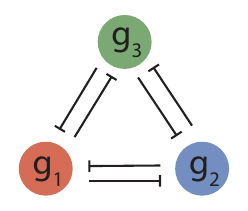}
    \caption{Network of three mutually-repressing genes. Transcribing each gene $g_{i}$ produces a repressor with average concentration denoted by $R_{i}$.}
    \label{fig:3genetogglediagram}
\end{figure}

The three-gene toggle switch, shown in
Figure~\ref{fig:3genetogglediagram}, provides a powerful basis for examining the rise of tristability and higher-order multistability in a range of biological contexts. The motif is most directly useful in a synthetic setting, where dynamic shifts can be tuned precisely through experiments. However, the model can also help uncover how various sets of three (or more) key genes interact to generate the multistable cell fates observed in nature. For example, experimental work has identified sets of three ``master regulator'' genes that interact to drive protein expression toward three or more distinct cell fate phenotypes, such as in T helper cell differentiation~\cite{TcellFangZhu:2017} and stem cell differentiation~\cite{OctSoxNanog:2020}.

Recent work has also begun to build a comprehensive picture of stable expression in such three-gene networks~\cite{CorsonSiggia:2012,DudduJolly:2020,DudduJolly:2022}. Our study focuses instead on alternative approaches to induction in a Hill model for the three-gene toggle. Within this reframed context for tuning dynamics, we then investigate not only how the mechanisms of induction shift transcription factor activities and stable state expression levels, but also how the dynamic topologies of the resulting fixed point landscapes evolve.

\begin{figure*}[t]
    \centering
    \includegraphics[width=\textwidth]{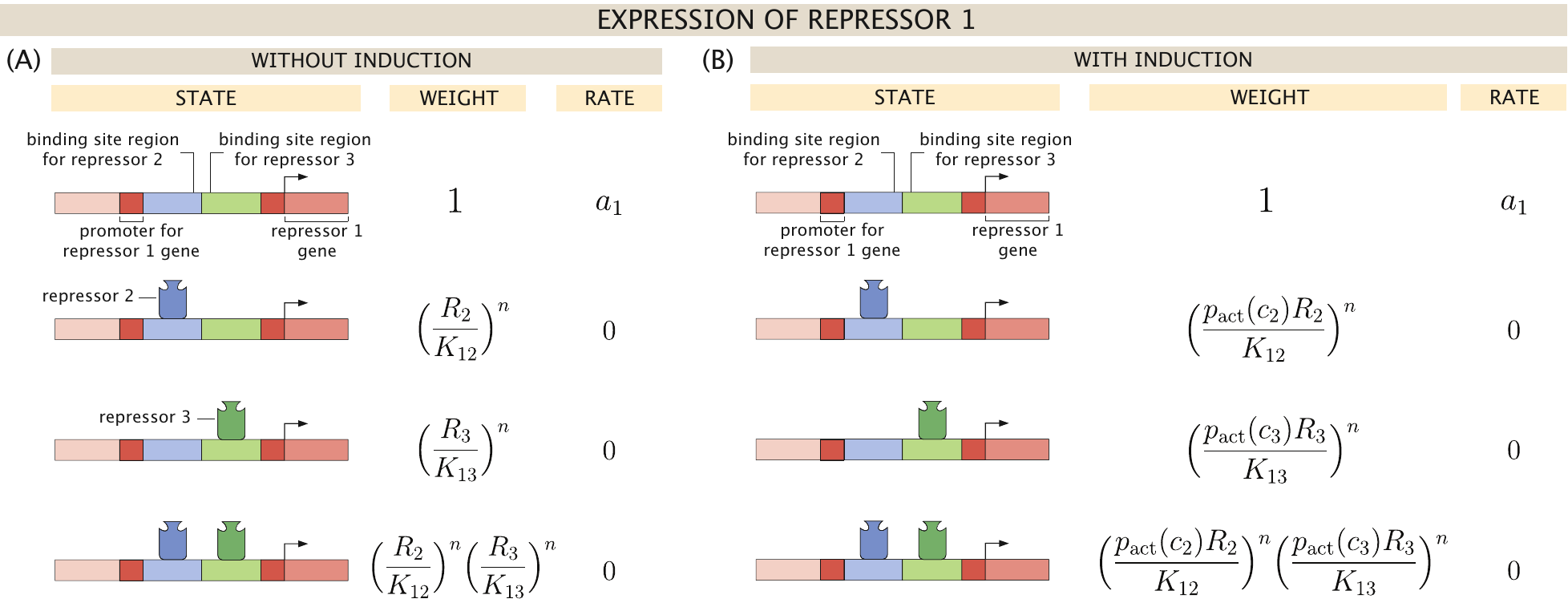}
    \caption{Expression of repressor 1 in the three-gene toggle switch, as described by a Hill function model. Equivalent definitions apply to expression of repressors 2 and 3. (A) Thermodynamic states, weights, and rates for expression of repressor 1 ($R_{1}$). The regulating repressors ($R_{2}$ and $R_{3}$) bind non-exclusively to the target promoter region to suppress gene transcription. Each repressor $R_{i}$ binds at the promoter region for repressor $R_{1}$ with affinity $K_{1i}$ and cooperativity $n$. (B) Thermodynamic states, weights, and rates for expression of repressor $R_{1}$ in the presence of inducers. Expression now depends on the \emph{active} concentration of each repressor $R_{i}$, which is determined by a distinct inducer at concentration $c_{i}$, defining the probability of activity $p_{\text{act}}(c_{i})$.}
    \label{fig:statesandweights_standard}
\end{figure*}

We begin in Section II by defining our model for the three-gene toggle switch, and showing how it mathematically incorporates the regulatory activity of transcription factors as allosterically-inducible proteins. From this initial framework, Section III demonstrates the different dynamics that become possible as inducer concentrations rise. In Section IV we highlight how the thresholds, i.e., bifurcations, that separate different dynamic regimes shift when perturbed away from an initial assumed symmetry in gene interactions. Section V explores the impact of self-activation on the complexity of dynamics and the number of stable steady states possible for our three-gene switch. In particular, we find that the biological interpretation of dual repression and self-activation, whether arising from binding exclusivity or the effector's role, can significantly affect the types of dynamics that unfold. In so doing, our work sheds light on the complexity of dynamics possible from simple modeling choices, and on just how precisely inducer concentrations must be tuned under different conditions for a system to access various dynamic regimes. The diverse range of input-output responses demonstrated across the different three-gene networks presented here motivates further study of how our results generalize to higher-order gene regulatory architectures.

\section{The three-gene toggle switch}
We first consider the most direct extension of the two-gene bistable toggle switch to three genes, as depicted in Fig.~\ref{fig:3genetogglediagram}. Transcribing a given gene $g_{i}$ generates a protein $R_{i}$ called a transcription factor, which can in turn bind to the promoter region of either remaining gene. In this particular form of the model, the transcription factor prevents RNA polymerase (RNAP) from binding and initiating transcription, thereby repressing expression. We begin by discussing how to model the dynamics of this system of repressors before demonstrating the role of induction in governing how dynamical behavior evolves as a function of inducer concentration(s).

\subsection{The baseline model}\label{subsec:baseline}

The modeling of stability in basic switches and other regulatory motifs has long been explored using the tools of dynamical systems~\cite{Goodwin1963,CherryAdler:2000,Covert2015,Alon2020,Ferrell2022}. To describe the expression patterns that arise from the three-gene toggle switch architecture shown in Fig.~\ref{fig:3genetogglediagram}, we define a dynamic equation for each transcription factor repressor $R_{i}$ of the form
\begin{equation}
    \frac{dR_{i}}{dt} = a_{i}p_{i}^{\text{expr}}(R_{i},\: \{R_{j\neq i}\}) - \frac{1}{\tau_{i}}R_{i}. \label{eq:Reqgenform}
\end{equation}
This general form assumes that the dynamics of gene transcription, i.e., the production of mRNA transcripts, evolves at a timescale such that production can be measured equivalently from the concentration of output repressor $R_{i}$. The first term of Eqn.~\ref{eq:Reqgenform} characterizes protein production at a maximal rate $a_{i}$, and depends on the probability $p_{i}^{\text{expr}}$ that expression can occur. The second term accounts for protein degradation at a rate $1/\tau_i$.

Fig.~\ref{fig:statesandweights_standard}(A) lists the possible regulatory states and corresponding statistical weights that contribute to the probability of expression $p_{i}^{\text{expr}}$. This approach to describing biological regulation draws from well-established statistical mechanical principles~\cite{TLHill1977,TLHill1989,Ackers1982,Shea1985,bintu2005transcriptional,bintu2005transcriptional2,Buchler2003a,Kuhlman2007,Vilar2003a,Vilar2003b,vilarSaiz2013}. From this table, and equivalent definitions for expression of repressors 2 and 3, we define the dynamics of the corresponding three-gene circuit by the set of differential equations
\begin{eqnarray}
    \frac{dR_{1}}{dt} &=& \frac{a_{1}}{\Big[1+\Big(\frac{R_{2}}{K_{12}}\Big)^n\Big]\Big[1+\Big(\frac{R_{3}}{K_{13}}\Big)^n\Big]} -\frac{1}{\tau_{1}}R_{1},\label{eq:ODEnoc1}\\
    \frac{dR_{2}}{dt} &=& \frac{a_{2}}{\Big[1+\Big(\frac{R_{1}}{K_{21}}\Big)^n\Big]\Big[1+\Big(\frac{R_{3}}{K_{23}}\Big)^n\Big]} -\frac{1}{\tau_{2}}R_{2},\label{eq:ODEnoc2}\\
    \frac{dR_{3}}{dt} &=& \frac{a_{3}}{\Big[1+\Big(\frac{R_{1}}{K_{31}}\Big)^n\Big]\Big[1+\Big(\frac{R_{2}}{K_{32}}\Big)^n\Big]} -\frac{1}{\tau_{3}}R_{3}.\label{eq:ODEnoc3}
\end{eqnarray}
Each repressor $R_j$ binds non-exclusively with affinity $K_{ij}$ to the promoter region for the gene that expresses repressor $R_i$. This repression coefficient $K_{ij}$ reflects the characteristic concentration of repressor $R_j$ required to strongly regulate $R_i$'s expression. In other words, when $R_j = K_{ij}$, the produced concentration of repressor $R_i$ is half its maximum possible value.

Note that these equations describe transcription factor binding phenomenologically through a Hill function model with coefficient $n$~\cite{Hill1910,Hill1913}. This approach coarse-grains out the molecular detail of precise binding site occupancies found in a strictly thermodynamic model, instead representing its high cooperativity limit. The Hill coefficient $n$ thus does not necessarily correspond to the number of bound repressor molecules, but rather measures the sensitivity of output response to binding. 

While there are advantages to the specificity of thermodynamic models~\cite{yangrousseaumahdavi2025}, using them to evaluate stability profiles quickly becomes intractable in higher-dimensional networks, including those analyzed here. We therefore take a Hill model approach for the remainder of the discussion. Even without full knowledge of precise binding site occupancies, the Hill function offers a strong fit to empirical data for gene regulatory dynamics~\cite{Gardner:2000,Winfree2006}. The model also reflects the Hopfield barrier for the sharpness of input-output response~\cite{GunawardenaHill2024}, meaning that for a given coefficient $n$, the Hill model reflects the strongest, most sigmoidal input-output response possible for a system without energy expenditure. The approach thus allows us to investigate dynamics under the most input-sensitive model conditions, and to observe its maximum possible dynamic complexity.

We now have a model for the toggle switch that depends on various theoretically accessible parameters, specifically dissociation constants $K_{ij}$, production rates $a_i$, and degradation rates $1/\tau_i$. As pointed out previously, however, these parameters as written are not easily accessed experimentally, and are disconnected from the underlying biology that controls them within living cells. While the current model indicates that regulation is controlled by the total concentrations of transcription factors, these proteins are often themselves controlled by the allosteric binding of effector molecules. This implies that cells respond to changes in the number of \emph{active} transcription factors. 

In the following subsections we will use statistical mechanics to outline the role of effector molecules (specifically inducers) in defining transcription factor activity.

\subsection{A statistical mechanical model for allosteric induction}\label{sec:allostery}

Shifting to an allosteric description of gene regulation, we now argue formally that gene expression depends not on the total concentration of the regulating transcription factor, $[\text{TF}_{\text{tot}}]$, but rather on the \emph{active} concentration thereof, i.e.,
\begin{equation}
    [\text{TF}_{\text{act}}] = p_{\text{act}}(c)[\text{TF}_{\text{tot}}],
\end{equation}
where $p_{\text{act}}(c)$ is the probability that the transcription factor is active as a function of effector concentration $c$. This probability can be defined by the Monod-Wyman-Changeux (MWC) model~\cite{Gerhart1962, Monod1963,Monod1965,o1980equilibrium,Marzen2013,Changeux2013,Gerhart2014,Phillips:2020}, which states that when an effector at concentration $c$ can bind allosterically at $m$ sites on a transcription factor, the probability of that transcription factor being active is
\begin{equation}
    p_{\text{act}}(c) = \frac{\Big(1+\frac{c}{K_A}\Big)^m}{\Big(1+\frac{c}{K_A}\Big)^m + e^{-\beta\Delta\epsilon}\Big(1+\frac{c}{K_I}\Big)^m},\label{eq:pact}
\end{equation}
where $K_A$ and $K_I$ are the dissociation constants for active and inactive repressor states, respectively, $\beta = 1/k_B T$, and $\Delta\epsilon = \epsilon_{I}-\epsilon_{A}$ is the energy difference between inactive ($\epsilon_{I}$) and active ($\epsilon_{A}$) states. Repressors are driven toward the inactive state upon effector binding when $K_I < K_A$. Effectors obeying this condition are referred to as inducers. Conversely, when $K_I > K_A$, repressors are driven toward the active state, and effectors act as corepressors. Examples of both phenomena have been experimentally isolated in nature~\cite{Jobe1972,Foster2010,Yanofsky1974}, but for simplicity we will assume the case of $K_I < K_A$ in what follows and refer exclusively to inducers.

\begin{figure}[t]
    \centering
    \includegraphics[width=\columnwidth]{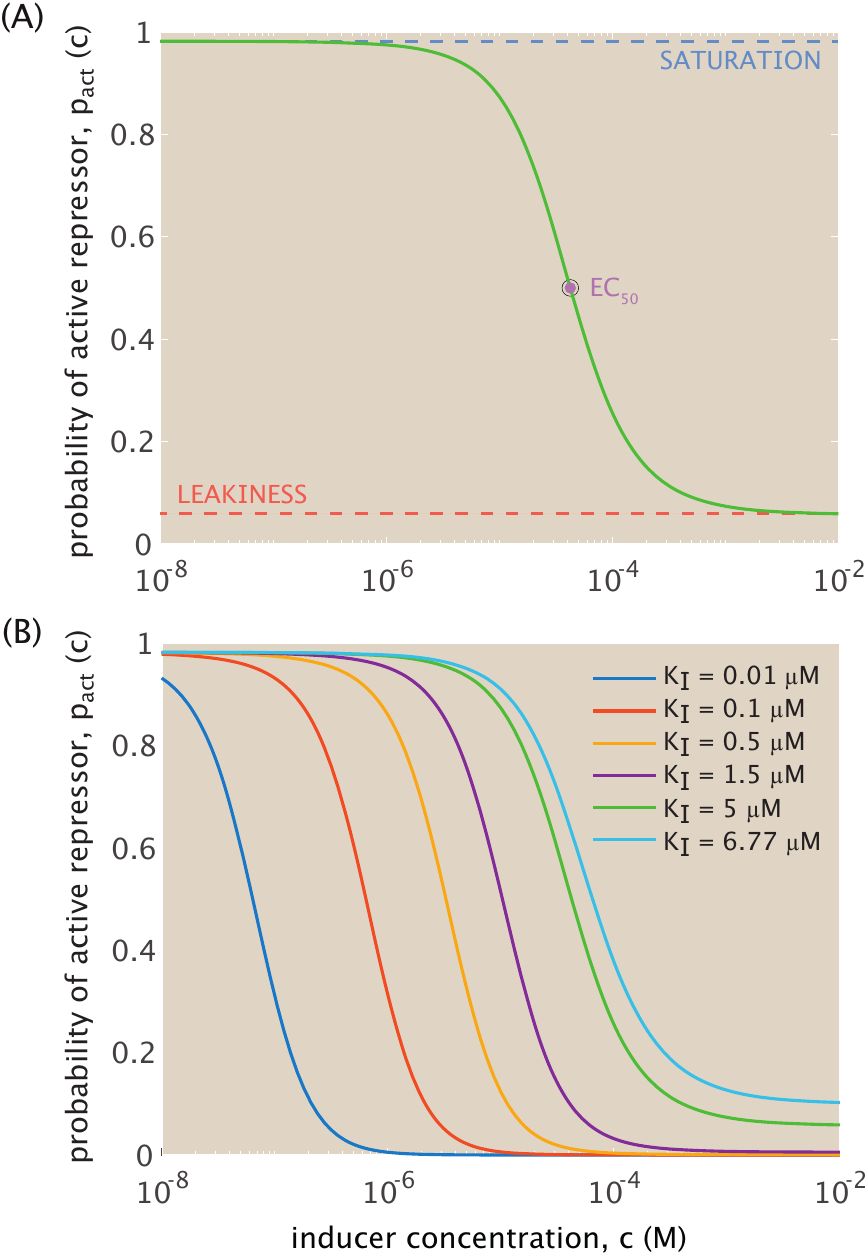}
    \caption{Repressor activity as a function of inducer concentration $c$, with $K_I < K_A$. (A) The probability of active repressor as a function of inducer concentration, with $m = 2$, $\Delta\epsilon = 4\:k_BT$, $K_A = 150$ $\mu M$, and $K_I = 5$ $\mu M$. The saturation and leakiness limits are denoted in blue and orange, respectively, and the $EC_{50}$, i.e., inducer concentration at which the probability is half maximal, is marked in purple. (B) Evolution of the probability for different values of $K_I$, with the curve in panel (A) shown again in green. The shift in probability spans the range of allowable $K_I$ values for the specified parameters up to the boundary at $K_I = 6.77\:\mu M$, derived from Appendix~\ref{app:MWC}.}
    \label{fig:prob}
\end{figure}

Fig.~\ref{fig:prob}(A) plots one possible probability curve $p_{\text{act}}(c)$ in the induction setting with $K_I < K_A$ for a fixed set of parameters. This function follows standard sigmoidal behavior, with a sufficient increase in inducer concentration rendering repressor inactive. Depending on how we define parameters, we can alter several key properties of the input-output response from this probability function. The leakiness represents the amount of activity at saturating concentrations of inducer, i.e., $p_{\text{act}}(c\rightarrow\infty)$, and the saturation conversely denotes the amount of activity in the absence of inducer ($p_{\text{act}}(c = 0)$). The inducer concentration at which the probability is half maximal is denoted as the $EC_{50}$ value, and the sharpness of the input-output response curve at this inflection point rises with an increasing Hill coefficient $n$. The MWC parameters chosen therefore determine how effectively inducers suppress the activity of repressors and the sensitivity of the output response to an increase in inducer concentration.

Fig.~\ref{fig:prob}(B) highlights several viable probability curves, defined by MWC parameters bounded to biologically permissible ranges derived from Eqns.~\ref{eq:boundsq01_1} and~\ref{eq:boundsq01_2} in Appendix~\ref{app:MWC}. We assume throughout that there are $m = 2$ binding sites for inducers on repressors. For fixed $\beta\Delta\epsilon = 4$ and $K_A = 150$ $\mu M$, Fig.~\ref{fig:prob}(B) then maps the probability of activity as a function of molar inducer concentration for increasing dissociation constant $K_I < K_A$. As the ratio $K_A/K_I$ decreases, approaching its lower bound (denoted in cyan as $K_I = 6.77$ $\mu M$ in Fig.~\ref{fig:prob}), the curve's inflection point shifts toward higher inducer concentrations. This occurs because as $K_I$ rises and the inducer becomes less tightly bound to the inactive compared to the active repressor, a higher inducer concentration is necessary to sequester inactive repressors equivalently. Given that the choice of parameters in Fig.~\ref{fig:prob}(A) is also within the biologically reasonable regime, we focus on these parameters for the remainder of our study, unless otherwise stated.

Note that increasing the number of inducer binding sites $m$ for the target repressor protein would increase the cooperativity of inducer-repressor interactions, and thus increase the sensitivity of response to the presence of inducer. This would make the slope of the probability function steeper at the inflection point. Keeping all other properties fixed, an increase in $m$ also decreases the inducer concentration required to suppress repressor activity.

While the role of induction has been widely discussed and used in experimental literature, theoretical studies leveraging this perspective for parametrization remain ongoing, with recent work demonstrating the utility of induction in modeling the dynamics of simple gene regulatory motifs~\cite{Razo2018,yangrousseaumahdavi2025}. This approach, however, has not yet been expanded to the more intricate regulatory networks that are standard in nature. Toward this end, we now incorporate the MWC description for the probability of active repressor into our model for gene expression dynamics in the three-gene toggle switch.

\subsection{The baseline model as a function of inducer concentrations}

Suppose now that each repressor produced from the three-gene toggle circuit in Fig.~\ref{fig:3genetogglediagram} can be controlled by its own inducer, each at concentrations $c_{1}$, $c_{2}$, and $c_{3}$. We will make the reasonable assumption that each repressor responds to induction with the same sensitivity and thus in accord with the same probability function. As shown in Fig.~\ref{fig:statesandweights_standard}(B), this means that expression of repressor $1$ depends on the concentrations of active repressors $2$ and $3$, namely, $p_{\text{act}}(c_2)\times R_{2}$ and $p_{\text{act}}(c_3)\times R_{3}$. We similarly scale the regulatory contribution of $R_1$ to expression of $R_2$ and $R_3$ by the probability $p_{\text{act}}(c_1)$. Eqns.~\ref{eq:ODEnoc1} -~\ref{eq:ODEnoc3} thus become
\begin{small}
\begin{eqnarray}
    \frac{dR_{1}}{dt} &=& \frac{a_{1}}{\Big[1+\Big(\frac{p_{\text{act}}(c_{2})R_{2}}{K_{12}}\Big)^n\Big]\Big[1+\Big(\frac{p_{\text{act}}(c_{3})R_{3}}{K_{13}}\Big)^n\Big]} -\frac{R_1}{\tau_{1}},\quad \label{eq:3Gr1induce}\\
    \frac{dR_{2}}{dt} &=& \frac{a_{2}}{\Big[1+\Big(\frac{p_{\text{act}}(c_{1})R_{1}}{K_{21}}\Big)^n\Big]\Big[1+\Big(\frac{p_{\text{act}}(c_{3})R_{3}}{K_{23}}\Big)^n\Big]} -\frac{R_2}{\tau_{2}},\quad\label{eq:3Gr2induce}\\
    \frac{dR_{3}}{dt} &=& \frac{a_{3}}{\Big[1+\Big(\frac{p_{\text{act}}(c_{1})R_{1}}{K_{31}}\Big)^n\Big]\Big[1+\Big(\frac{p_{\text{act}}(c_{2})R_{2}}{K_{32}}\Big)^n\Big]} -\frac{R_3}{\tau_{3}}. \label{eq:3Gr3induce}
\end{eqnarray}
\end{small}
We can simplify the model to a dimensionless form through several assumptions. First, we assume for analytic convenience that all repressors are produced and degraded at the same maximal rates, independent of the underlying dynamics of transcription. This sets $a_1 = a_2 = a_3 \equiv a$ and $\tau_1 = \tau_2 = \tau_3 \equiv \tau$. We also assume that a given repressor binds to different promoter sites with the same affinity, such that $K_{21} = K_{31} \equiv K_1$, $K_{12} = K_{32} \equiv K_2$, and $K_{13} = K_{23} \equiv K_3$. This allows us to re-express Eqns.~\ref{eq:3Gr1induce} -~\ref{eq:3Gr3induce} in dimensionless form, transforming $\bar{t} = t/\tau$, $\bar{R}_{i} = R_i/K_{i}$, and $\bar{a}_{i} = \tau a/K_{i}$ to obtain
\begin{small}
\begin{eqnarray}
    \frac{d\bar{R}_{1}}{d\bar{t}} &=& \frac{\bar{a}_1}{[1+(p_{\text{act}}(c_{2})\bar{R}_{2})^n][1+(p_{\text{act}}(c_{3})\bar{R}_{3})^n]}-\bar{R}_{1},\label{eq:ODEdimless_r1pre}\\
    \frac{d\bar{R}_{2}}{d\bar{t}} &=& \frac{\bar{a}_2}{[1+(p_{\text{act}}(c_{1})\bar{R}_{1})^n][1+(p_{\text{act}}(c_{3})\bar{R}_{3})^n]}-\bar{R}_{2},\label{eq:ODEdimless_r2pre}\\
    \frac{d\bar{R}_{3}}{d\bar{t}} &=& \frac{\bar{a}_3}{[1+(p_{\text{act}}(c_{1})\bar{R}_{1})^n][1+(p_{\text{act}}(c_{2})\bar{R}_{2})^n]}-\bar{R}_{3}. \label{eq:ODEdimless_r3pre}
\end{eqnarray}
\end{small}
Finally, for most of the discussion we consider a symmetric system in which all repressors bind at all sites with equal affinity such that $K_1 = K_2 = K_3 \equiv K$. (We will examine how breaking this symmetry affects shifts in dynamics in later sections.) This means that $\bar{a}_1 = \bar{a}_2 = \bar{a}_3 = \bar{a}$, and we arrive at the system of equations
\begin{small}
\begin{eqnarray}
    \frac{d\bar{R}_{1}}{d\bar{t}} &=& \frac{\bar{a}}{[1+(p_{\text{act}}(c_{2})\bar{R}_{2})^n][1+(p_{\text{act}}(c_{3})\bar{R}_{3})^n]}-\bar{R}_{1},\label{eq:ODEdimless_r1}\\
    \frac{d\bar{R}_{2}}{d\bar{t}} &=& \frac{\bar{a}}{[1+(p_{\text{act}}(c_{1})\bar{R}_{1})^n][1+(p_{\text{act}}(c_{3})\bar{R}_{3})^n]}-\bar{R}_{2},\label{eq:ODEdimless_r2}\\
    \frac{d\bar{R}_{3}}{d\bar{t}} &=& \frac{\bar{a}}{[1+(p_{\text{act}}(c_{1})\bar{R}_{1})^n][1+(p_{\text{act}}(c_{2})\bar{R}_{2})^n]}-\bar{R}_{3}. \label{eq:ODEdimless_r3}
\end{eqnarray}
\end{small}
\noindent The original model in Eqns.~\ref{eq:ODEnoc1} -~\ref{eq:ODEnoc3} thus collapses from a high-dimensional parameter space to at most a three-dimensional space defined by inducer concentrations $c_1$, $c_2$, and $c_3$, which can be directly controlled experimentally.  Note that while the symmetry introduced here is not necessary, it makes the underlying dynamics more transparent.

\section{Regimes of multistability}

\begin{figure*}
    \centering
    \includegraphics[width=\textwidth]{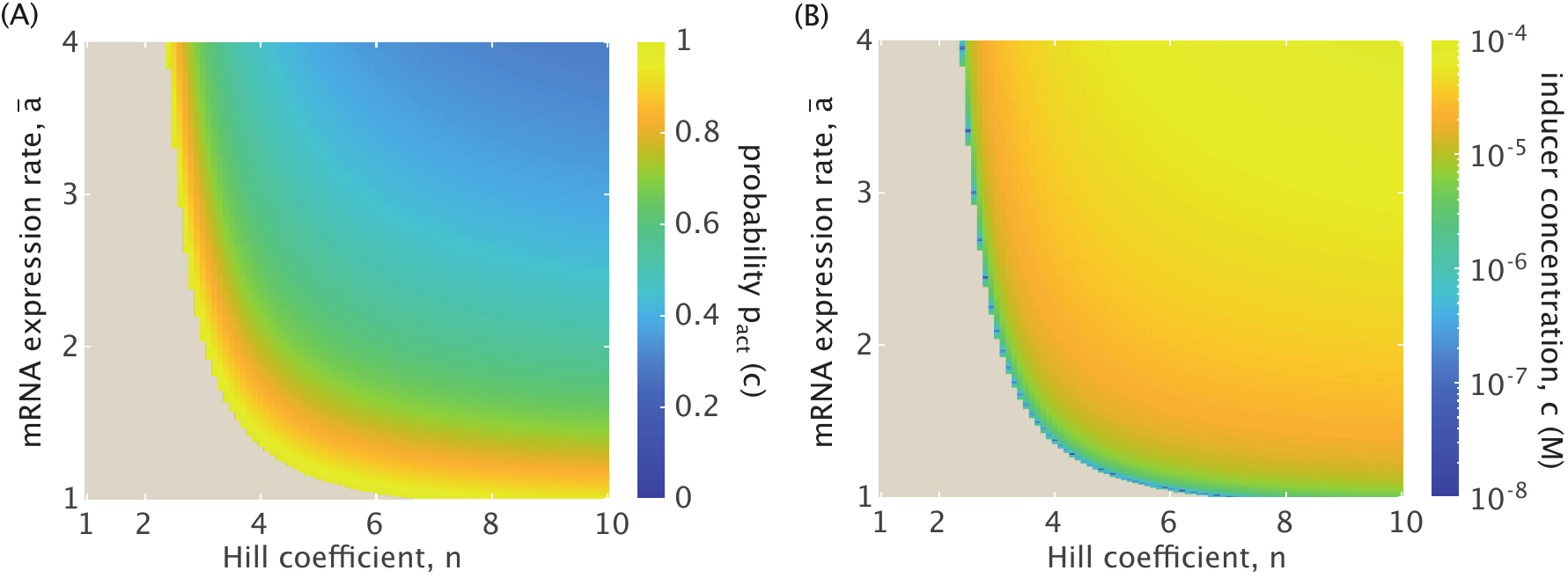}
    \caption{Heatmaps tracking the bifurcation threshold beyond which the three-gene toggle switch can no longer have an $\bar{R}_1$-dominant steady state and be tristable. Note that we define the limit steady state case here as satisfying $\bar{R}_2^n$, $\bar{R}_3^n\rightarrow 0$, with $\bar{R}_2 = \bar{R}_3 \leq \varepsilon$. The results shown are specific to the choice $\varepsilon = \sqrt[n]{0.01}$. (A) The threshold probability of active repressor as a function of Hill coefficient $n$ and mRNA expression rate $\bar{a}$. Note that the beige-shaded region for smaller axis values denotes a regime in which the steady state is never possible at any inducer concentration. (B) The threshold specified in Fig.~\ref{fig:prob}(A), now in terms of inducer concentration for the MWC model.}
    \label{fig:OG_bistabthresh}
\end{figure*}

For a given Hill exponent $n$ and maximal (dimensionless) production rate $\bar{a}$, we now explore the dynamic profiles possible for the three-gene toggle switch as a function of inducer concentrations. In particular, we quantify the thresholds, or bifurcations, as a function of inducer that bring about fundamental shifts in dynamics. Our graphical analysis in this section will consider the large cooperativity case where $n = 4$, along with setting $\bar{a} = 2$. While we will also examine how much the dynamic thresholds shift as we alter parameters, our focus in this section is specifically on the effects of induction at various intensities. To this end, we begin by considering a system tuned by a single inducer.

\subsection{Single inducer}\label{sec:inducer1}

Suppose a single inducer controls the dynamics of the three-gene toggle switch by regulating the activity of $R_1$. This means that $p_{\text{act}}(c_1) \equiv p_{\text{act}}(c)$ and $p_{\text{act}}(c_{2}) = p_{\text{act}}(c_3) = 1$ such that $\bar{R}_2$ and $\bar{R}_3$ are always maximally active. Substituting these definitions into Eqns.~\ref{eq:ODEdimless_r1} -~\ref{eq:ODEdimless_r3}, we obtain the single-inducer model
\begin{align}
    \frac{d\bar{R}_{1}}{d\bar{t}} &= \frac{\bar{a}}{(1+\bar{R}_{2}^n)(1+\bar{R}_{3}^n)}-\bar{R}_{1},\label{eq:inducer1_r1}\\
    \frac{d\bar{R}_{2}}{d\bar{t}} &= \frac{\bar{a}}{[1+(p_{\text{act}}(c)\bar{R}_{1})^n](1+\bar{R}_{3}^n)}-\bar{R}_{2},\label{eq:inducer1_r2}\\
    \frac{d\bar{R}_{3}}{d\bar{t}} &= \frac{\bar{a}}{[1+(p_{\text{act}}(c)\bar{R}_{1})^n](1+\bar{R}_{2}^n)}-\bar{R}_{3}. \label{eq:inducer1_r3}
\end{align}
For a given inducer concentration $c$, one then sets the above equations to zero and solves to obtain the steady state expressions.

We can first gain some analytical insight from the model by considering certain limit cases, specifically those regarding the existence of single-repressor-dominant steady states, as a function of inducer concentration. Consider the existence of an $\bar{R}_2$-dominant steady state whereby $\bar{R}_1 = \bar{R}_3 = \varepsilon$ such that $\bar{R}_1^n,\:\bar{R}_3^n\rightarrow 0$. It follows from setting Eqn.~\ref{eq:inducer1_r2} to zero that such a steady-state requires $\bar{R}_2 = \bar{a}$. By similarly setting Eqns.~\ref{eq:inducer1_r1} and~\ref{eq:inducer1_r3} to zero, it also follows that $\bar{R}_{1,3} = \bar{a}/(1+\bar{R}_2^n)$. We therefore conclude that the system always has the following steady state regardless of inducer concentration:
\begin{equation}
    \bar{R}_{ss} = (\bar{R}_1,\:\bar{R}_2,\:\bar{R}_3) = \Big(\frac{\bar{a}}{1+\bar{a}^n},\bar{a},\frac{\bar{a}}{1+\bar{a}^n}\Big). \label{eq:R2dom}
\end{equation}
Similar logic for the existence of an $\bar{R}_3$-dominant state leads to a corresponding predicted steady state
\begin{equation}
    \bar{R}_{ss} = (\bar{R}_1,\:\bar{R}_2,\:\bar{R}_3) = \Big(\frac{\bar{a}}{1+\bar{a}^n},\frac{\bar{a}}{1+\bar{a}^n},\bar{a}\Big). \label{eq:R3dom}
\end{equation}

The conditions for an $\bar{R}_1$-dominant steady state, where $\bar{R}_2^n,\:\bar{R}_3^n\rightarrow 0$, follow from the same logic in Eqns.~\ref{eq:inducer1_r1} -~\ref{eq:inducer1_r3} as $\bar{R}_1 = \bar{a}$ and
\begin{equation}
    \bar{R}_2 = \bar{R}_3 = \frac{\bar{a}}{1+(p_{\text{act}}(c)\bar{R}_1)^n}.\label{eq:R1dom}
\end{equation}
Unlike the $\bar{R}_2$ and $\bar{R}_3$-dominant states in Eqns.~\ref{eq:R2dom} and~\ref{eq:R3dom}, however, Eqn.~\ref{eq:R1dom} indicates that for the $\bar{R}_1$-dominant steady state the concentrations of non-induced repressors $\bar{R}_2$ and $\bar{R}_3$ depend on inducer concentration $c$. As $c$ increases and the probability of activity decreases, the corresponding repressor concentrations at steady state rise. This also means that at a sufficiently high inducer concentration, $\bar{R}_2^n,\:\bar{R}_3^n\rightarrow 0$ no longer holds and the steady state can no longer exist. Assuming $\bar{R}_2 = \bar{R}_3 \leq \epsilon$ satisfies the specified condition, we derive the inducer bifurcation threshold by substituting this inequality into Eqn.~\ref{eq:R1dom} and rearranging to obtain the result
\begin{equation}
    p_{\text{act}}(c) \geq \frac{1}{\bar{a}}\sqrt[n]{\frac{\bar{a}}{\varepsilon}-1}.\label{eq:R1thresh}
\end{equation}
At equality, the probability in Eqn.~\ref{eq:R1thresh} represents a bifurcation threshold, where for lower probabilities (i.e, higher inducer concentrations) the system can no longer exist at an $\bar{R}_1$-dominant steady state.

\begin{figure*}[t]
    \centering
    \includegraphics[width=\textwidth]{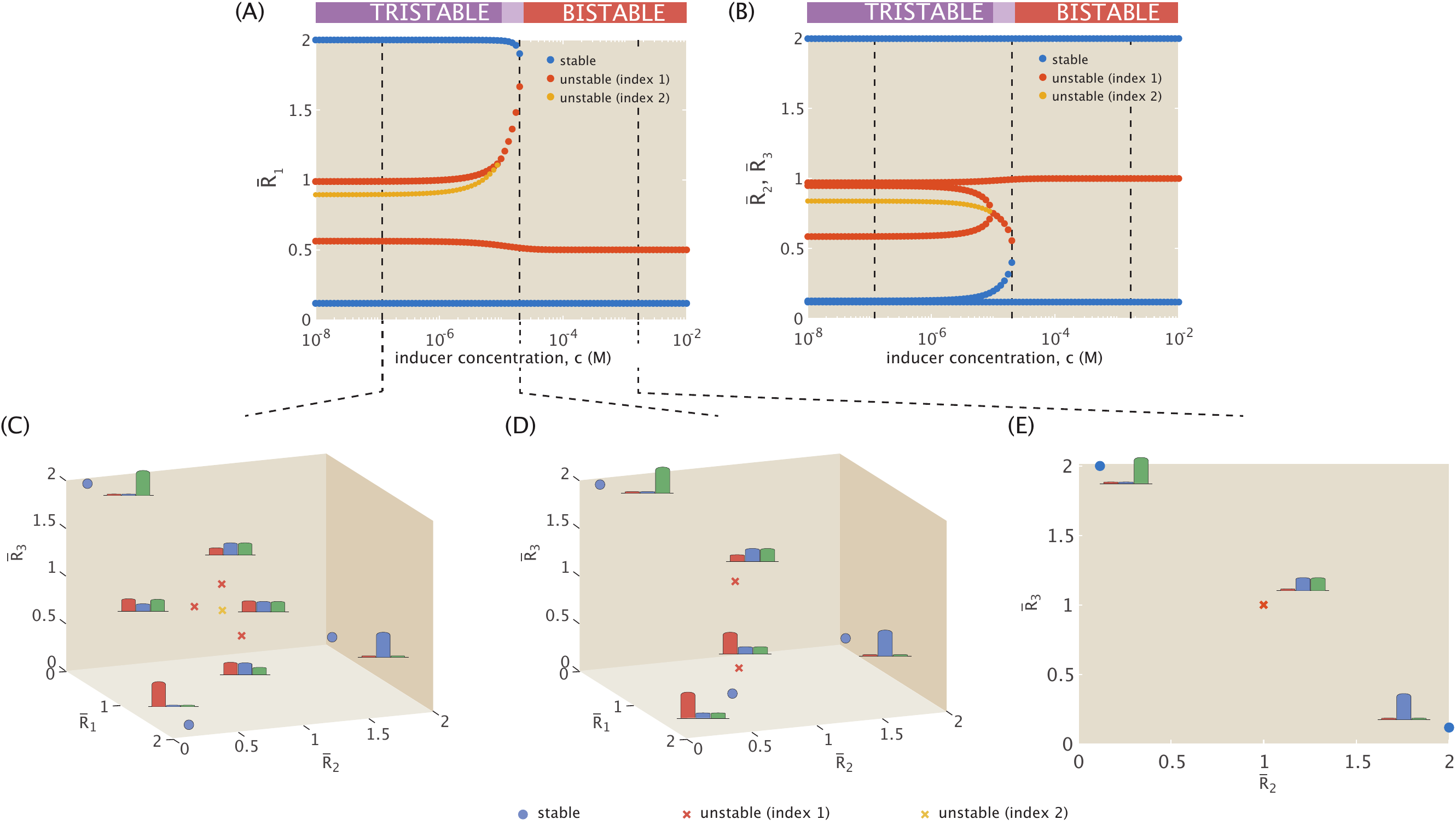}
    \caption{Dynamics of the three-gene toggle switch with $n = 4$ and $\bar{a} = 2$ for increasing inducer concentration $c$. Allosteric regulation is defined by the probability curve shown in Fig.~\ref{fig:prob}(A). (A)-(B) Bifurcation diagrams for $\bar{R}_1$, $\bar{R}_2$, and $\bar{R}_3$ steady state expression as a function of inducer concentration. (C) Fixed points for the three-gene toggle switch at a low inducer concentration. (D) Fixed points for the three-gene toggle switch at an intermediate inducer concentration. (E) Fixed points for the three-gene toggle switch at a high inducer concentration. In (C)-(E) the stability of each fixed point is color-coded as in panels (A) and (B), and each fixed point is labeled with its corresponding expression levels, colored as in Fig.~\ref{fig:3genetogglediagram}.}
    \label{fig:inducer1}
\end{figure*}

Eqns.~\ref{eq:R2dom} -~\ref{eq:R1thresh} therefore indicate that at low inducer concentration the system can arrive at a steady state in which any gene among the three can dominate expression. When the probability of activity drops below the threshold in Eqn.~\ref{eq:R1thresh}, however, the presence of inducer is sufficient to suppress $\bar{R}_1$ activity, and the system can no longer maintain a steady state with $\bar{R}_1$-dominant expression. The sensitivity of $\bar{R}_1$ activity to tuning then depends on cooperativity (as approximated by Hill coefficient $n$) and on the protein expression rate $\bar{a}$.

Fig.~\ref{fig:OG_bistabthresh}(A) plots the probability threshold in Eqn.~\ref{eq:R1thresh} as a function of $n$ and $\bar{a}$. Note that the beige region of this heatmap depicts a regime where an $\bar{R}_1$-dominant steady state is not possible at any inducer concentration. When $\bar{a}\gtrsim 2$, increasing the expression rate has little effect on this cooperativity bound. Fig.~\ref{fig:OG_bistabthresh}(B) plots the same relationship between the parameters, now explicitly as a function of inducer concentration (from the probability definition of Eqn.~\ref{eq:pact} in Section~\ref{sec:allostery}). 

Fig.~\ref{fig:OG_bistabthresh} tracks cooperativities up to $n = 10$. While it is not common for known multimeric transcription factors to extend beyond tetramers, plotting higher $n$ acknowledges the uncertainty that remains in the field regarding the cooperative mechanics of eukaryotic regulation and particularly of enhancers~\cite{Panigrahi2021enhanc}. Indeed, there are emerging cases of eukaryotic transcription factors functioning as higher-order homomultimeric complexes, such as FOXP3 in regulatory T-cells~\cite{Zhang2023FOX}. It is therefore useful to observe whether such higher-order cooperativities would influence bifurcations in the system.

We find that the bifurcation threshold for the shift between tristable and bistable dynamics is most sensitive to parameter changes at the onset of the expression domain (i.e., at minimum required $n$ and $\bar{a}$). For $n \gtrsim 4$ and $\bar{a} \gtrsim 2$, both parameters are sufficiently large that increasing either has little effect on the threshold probability and corresponding inducer concentration. Given also the existence of homotetrameric transcription factors such as p53 that are crucial to cell fate decisions~\cite{McLure1998p53,Ly2020p53,Nicolini2023p53}, it is therefore reasonable to choose $n = 4$ as representative of dynamics in the highly-cooperative regime, and we do this for the remainder of the paper.

To show more completely how the dynamics evolve with increasing inducer concentration, we now apply the specific parameters $n = 4$ and $\bar{a} = 2$ to Eqns.~\ref{eq:inducer1_r1} -~\ref{eq:inducer1_r3}. At each inducer concentration, we numerically solve for the fixed point repressor concentrations. We then perform a linear stability analysis on each fixed point. We do this by evaluating the Jacobian which, for our three-gene network modeled by differential equations of the form $d\bar{R}_i/dt = f_i(\bar{R}_1,\bar{R}_2,\bar{R}_3)$, is a $3\times 3$ matrix with entries
\begin{equation}
    J_{ij} = \frac{\partial f_i}{\partial \bar{R}_j}.
\end{equation}
The Jacobian provides a first-order linear approximation of behavior evaluated near a fixed point, with its eigenvectors specifying the primary directions of dynamical motion from the fixed point, and with its eigenvalues reflecting the rates at which perturbations to the fixed point grow or decay along the corresponding eigenvectors. If all eigenvalues have negative real parts, perturbation in any eigenvector direction of expression space will decay back to the fixed point, indicating it to be stable. If any of the eigenvalues have a positive real part, however, perturbation in the corresponding eigenvector direction(s) will grow exponentially and the system will move away from the fixed point, rendering the fixed point unstable.

For the three-gene models we examine, the unstable fixed points are characterized as saddle points. Mathematically, this distinction arises because at least one eigenvalue is negative. Dynamic trajectories local to the fixed point thus move away in some dimensional directions (positive eigenvalues) and approach in others (negative eigenvalues). The number of positive eigenvalues determines the number of dimensional directions of ``escape'' from the fixed point and thus the ``index'' of the saddle. If we consider the dynamics of gene expression within a potential landscape, expression levels local to the highest index saddle essentially follow a hierarchical dynamic flow through expression space toward stable states, guided by the presence of saddle points with decreasing index.

Fig.~\ref{fig:inducer1} plots the resulting changes in the number of fixed points and their expression levels through bifurcation diagrams, isolating the fixed point trajectories of the $\bar{R}_1$ coordinate in panel (A), and the (overlapping) $\bar{R}_2$ and $\bar{R}_3$ coordinates in panel (B). Three types of fixed points emerge from analysis of the Jacobian. Stable equilibria are denoted in blue, unstable index-1 saddle points in red, and unstable index-2 saddles in yellow.

As the inducer concentration increases, the system shifts across three regimes of decreasing dynamic complexity. Fig.~\ref{fig:inducer1}(C)-(E) plot fixed points in expression space for a given inducer concentration within each of these regimes. At low inducer concentration, as highlighted in Fig.~\ref{fig:inducer1}(C), the system is tristable, stabilizing to solely favor expression of one of the three repressors. A sufficiently stochastic system can be perturbed to transition between any pair of these stable points through one of the index-1 saddle points. The system also has an index-2 saddle point at which all repressors are half-maximally expressed. This means that if a system were to have a sufficient concentration of each repressor and initially exist near this index-2 saddle, it would be equally likely for the system to settle at any of the three possible stable states.

At an intermediate inducer concentration, a pair of fixed points annihilate each other and a saddle bifurcation occurs. The resulting intermediate regime, represented in Fig.~\ref{fig:inducer1}(D), is still tristable, but the three stable states are now only connected by two saddle points. From an initial state with sufficient concentrations of all three repressors, the system can now evolve dynamically in one of two ways: by either falling into a bistable potential well favoring $\bar{R}_2$ or $\bar{R}_3$ expression, or by following a trajectory directly stabilizing toward $\bar{R}_1$-dominant expression. Dynamics are therefore more restricted in this regime than in the low inducer regime, and no longer equally likely to evolve toward any given stable state. Finally, Fig.~\ref{fig:inducer1}(E) shows fixed point expression levels at a high inducer concentration, where the presence of inducer sufficiently suppresses the activity of $\bar{R}_1$ and the system undergoes a saddle bifurcation to collapse down to a bistable switch between $\bar{R}_2$ and $\bar{R}_3$-dominant expression.

Fig.~\ref{fig:inducer1} thus allows us to determine the inducer concentration thresholds at which dynamic shifts occur, including the shift between tristable and bistable dynamics. Given a distribution of stable state expression levels from data, we could then distinguish between systems in the low inducer regime or in the more dynamically restricted intermediate regime. Fig.~\ref{fig:inducer1}(B) also highlights that it is only at intermediate inducer concentrations that the system does not necessarily stabilize to a state exclusively favoring a single gene's expression, and that it is this characteristic that uniquely distinguishes the low vs. intermediate inducer concentration regimes of tristability.

\begin{figure}[t]
    \centering
    \includegraphics[width=\columnwidth]{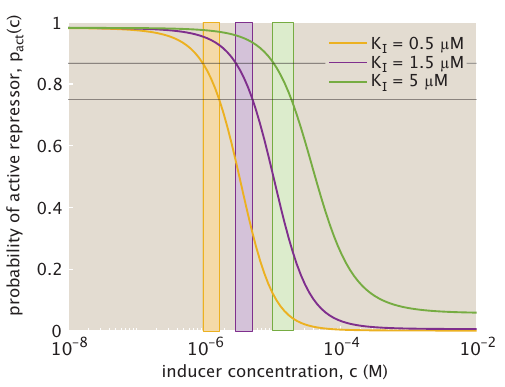}
    \caption{Change in inducer concentration range (shaded regions) of the intermediate dynamic regime for increasing dissociation constant $K_I$, overlaid with the corresponding activity probability curves from Fig.~\ref{fig:prob}. Note that regardless of $K_I$ parametrization, the intermediate regime is found within the same range of probabilities $0.749 \leq p_{\text{act}}(c) \leq 0.867$ (bounded in grey).}
    \label{fig:5fpchange}
\end{figure}

Given the relatively narrow concentration range for the intermediate dynamic regime, it is natural to ask how sensitive this window is to changes in allosteric interaction and to changes in the cooperativity, or ultrasensitivity, of expression response to induction. Regarding sensitivity to allostery, altering the ratio of $K_A$ and $K_I$ for an inducer binding to a repressor shifts the inflection point of the probability curve for activity toward higher inducer concentrations, as previously seen in Fig.~\ref{fig:prob}. We thus observe in Fig.~\ref{fig:5fpchange} that the log scale range of inducer concentrations within the intermediate dynamic regime does not change but shifts toward increasing value ranges with increasing $K_I$. Since altering the ratio of the dissociation constants does not affect the slope of the probability curve at the inflection point, the intermediate dynamic regime (for otherwise fixed values of $m = 2$ and $\beta\Delta\epsilon = 4$) exists across different $K_A/K_I$ for the same probability range, $0.749 \leq p_{\text{act}}(c) \leq 0.867$.

\begin{figure}[t]
    \centering
    \includegraphics[width=\columnwidth]{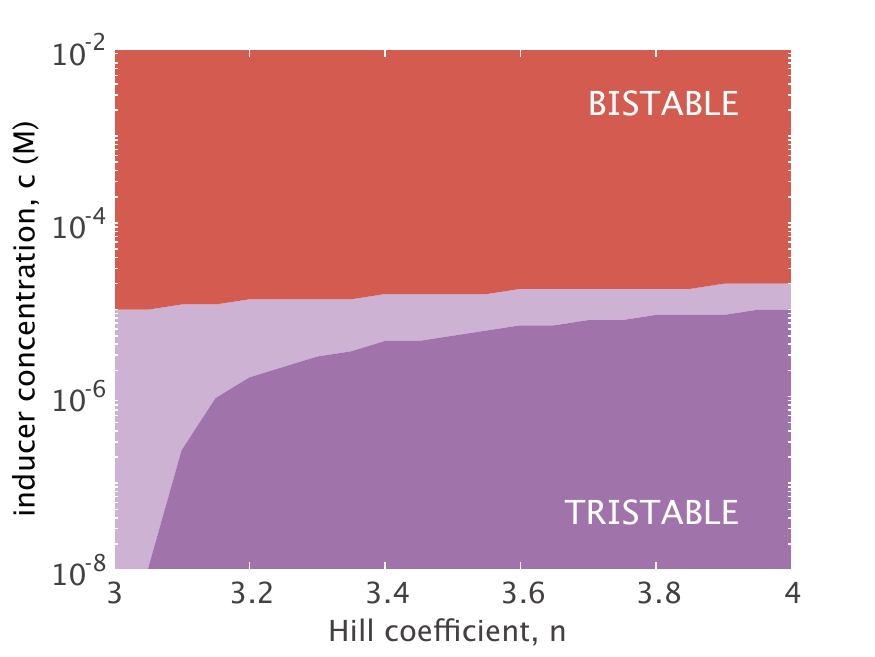}
    \caption{Change in the regions of multistability for $n \in [3, 4]$. Each dynamic region is color-coded as denoted in Fig.~\ref{fig:inducer1}, with the seven-fixed point tristable regime (purple), five-fixed point tristable regime (light purple), and three-fixed point bistable regime (red). For $n \lesssim 3$, the light purple region illustrates the most complex dynamic regime that is accessible.}
    \label{fig:5fpchangen}
\end{figure}

The size of the intermediate regime differs, however, with changing cooperativity. Fig. ~\ref{fig:OG_bistabthresh}(B) indicates that at $\bar{a} = 2$, bistability is possible for $n > 3$, but the threshold for shifting from tristability to bistability shifts toward lower inducer concentrations for Hill coefficients decreasing from $n < 4$. It is thus unclear from Fig.~\ref{fig:OG_bistabthresh} alone whether the range of inducer concentrations allowing the intermediate dynamic regime would expand, contract, or remain the same for a smaller cooperative Hill coefficient. Indeed, Fig.~\ref{fig:5fpchangen} demonstrates how the regimes of multistability evolve in response to varying cooperativity. As the Hill coefficient increases from $n = 3$, the inducer concentration at which the system bifurcates from tristable to bistable dynamics rises slightly in the log scale. The threshold separating the two distinct regimes of tristability, however, does significantly change. The inducer concentration range corresponding to the intermediate regime narrows with increasing cooperativity, since this implies a more sensitive response in gene expression to repressor binding, one that emphasizes the bias away from $\bar{R}_1$ expression as induction increases. The tristable threshold also begins to approach an apparent asymptote at $n = 4$, confirming our earlier finding that increasing cooperativity to $n = 4$ has little effect on the dynamics.

\subsection{Two inducers}\label{sec:2ind}

Activity in gene regulatory networks is not always restricted to control from a single effector. Our framework allows for induction by multiple effectors, whether synthetically through additional experimentally-inputted non-metabolizable inducers~\cite{Marbach2012}, or through naturally-occurring coordinated inducer activity. The direct involvement of small molecule effectors can be important, for example, in sugar-inducible genetic switches as they respond to metabolic nutrient availability ~\cite{Santillan2004, Hendrickson1984,Goodrich1992, SwintKruse2009, Shimizu2014}. Much remains to be learned, however, about how natural allosteric effectors coordinate cell fate differentiation~\cite{HuangEnver:2007}, and how existing regulators may play a role in inhibitory cross-talk, as evidenced in work on steroid hormone receptors~\cite{Weikum2017}. These possibilities make our study of three-gene toggle switches in the presence of two inducers even more relevant.

\begin{figure}[t]
    \centering
    \includegraphics[width=\columnwidth]{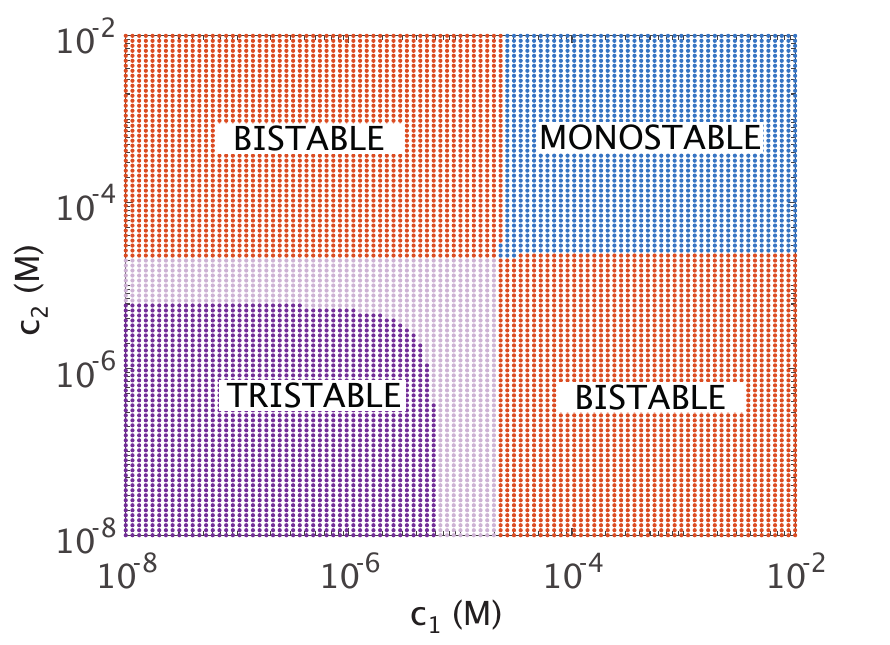}
    \caption{Phase diagram for evolving inducer concentrations $c_{1}$ and $c_{2}$ (regulating activity of $\bar{R}_1$ and $\bar{R}_2$, respectively). Each color-coded region corresponds to a phase defined by a different number of fixed points (seven in dark purple, five in light purple, three in red, and one in blue). Both inducers follow Eqn.~\ref{eq:pact} for the probability of activity with the same fixed parameters $m = 2$, $\Delta\epsilon = 4\:k_BT$, $K_A = 150$ $\mu M$, and $K_I = 5$ $\mu M$.}
    \label{fig:inducer2}
\end{figure}

We specifically extend to a system in which two repressors $\bar{R}_1$ and $\bar{R}_2$ are controlled by inducer concentrations $c_{1}$ and $c_{2}$, such that dynamics evolve by the differential equations
\begin{align}
    \frac{d\bar{R}_{1}}{d\bar{t}} &= \frac{\bar{a}}{(1+p_{\text{act}}(c_2)\bar{R}_{2}^n)(1+\bar{R}_{3}^n)}-\bar{R}_{1},\label{eq:inducer2_r1}\\
    \frac{d\bar{R}_{2}}{d\bar{t}} &= \frac{\bar{a}}{[1+(p_{\text{act}}(c_1)\bar{R}_{1})^n](1+\bar{R}_{3}^n)}-\bar{R}_{2},\label{eq:inducer2_r2}\\
    \frac{d\bar{R}_{3}}{d\bar{t}} &= \frac{\bar{a}}{[1+(p_{\text{act}}(c_1)\bar{R}_{1})^n](1+p_{\text{act}}(c_2)\bar{R}_{2}^n)}-\bar{R}_{3}. \label{eq:inducer2_r3}
\end{align}
Fig.~\ref{fig:inducer2} plots the shifts in dynamic regions of phase space (color-coded by the number of fixed points) for inducer concentrations $c_{1}$ and $c_{2}$. Sufficiently high concentrations of both inducers render the target repressors largely inactive, such that the system enters a monostable regime (plotted in blue) that stabilizes to a single point reflecting expression of only the remaining non-induced repressor.

\begin{figure*}[t]
    \centering
    \includegraphics[width=\textwidth]{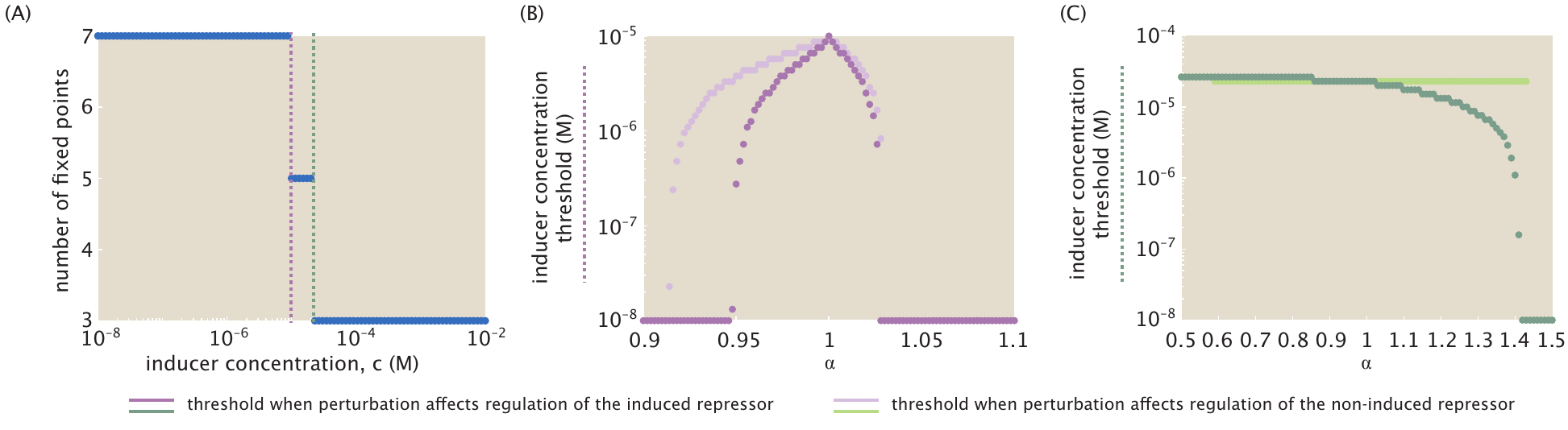}
    \caption{The change in bifurcation thresholds in a single inducer system as $\alpha$ rises. (A) The number of fixed points as a function of inducer concentration when $\alpha = 1$. The purple threshold denotes the transition between distinct tristable dynamic regimes (i.e., seven or five fixed points), and the green threshold denotes the transition between tristable and bistable dynamics (five or three fixed points, respectively). The system corresponds to that analyzed in Section ~\ref{sec:inducer1}. (B) The change in the inducer concentration threshold at which the tristable (purple) transition occurs as a function of $\alpha$. The dark purple curve corresponds to a system where $\alpha$ targets regulation of the induced repressor, $\bar{R}_1$. The light purple curve is for a system where $\alpha$ affects regulation of a non-induced repressor, in this case $\bar{R}_2$. (C) The change in the inducer concentration defining the threshold separating tristable from bistable dynamics as a function of $\alpha$. The specific colors again correspond to the different targets for perturbation bias, with dark green targeting the induced repressor and light green the non-induced repressor.}
    \label{fig:alphathreshchange}
\end{figure*}

The bifurcation line seen earlier at $p_{A}(c) \approx 0.725$ in Fig.~\ref{fig:inducer1} now becomes a set of two linear thresholds in two dimensions, intersecting at approximate inducer concentration $2.3\times 10^{-5}\:M$ (i.e., the approximate probability $p_{A}(c_{1}) = p_{A}(c_{2}) = 0.7$). This intersection defines four regions with one (monostability in blue), two (bistability in red), or three stable points (tristability in purple). Note that analytically we can show that the thresholds separating tristable from bistable dynamics for $c_1$ and $c_2$ are equal to those in Eqn.~\ref{eq:R1thresh} (see Appendix~\ref{sec:2indanalytic}). In the purple region of Fig.~\ref{fig:inducer2}, all three stable points become most accessible as both induction probabilities approach $1$. Note that when the probability of active repressor reaches $1$ for one of the repressors, the transition points along the axis of the other repressor reflect the bifurcations for the single-inducer system of Fig.~\ref{fig:inducer1}, as expected. Increasing $K_I$ for one of the inducers would shift the curve so that the intersection of its asymptote with the corresponding inducer axis occurs at a higher concentration. This trend also matches what we expect from the single inducer analysis.

The results observed for one and two inducers extend naturally in the three inducer regime, with tristability when all inducer concentrations are sufficiently small, monostability when all are sufficiently large, and bistability elsewhere. Similarly to how bifurcations transformed into a linear curve in two-dimensional phase space when moving from one to two regulating inducers, as in Fig.~\ref{fig:inducer2}, the three inducer case likewise extends these curves into corresponding bifurcation planes in three dimensions.

\section{Effects of induction in systems perturbed away from symmetry}

So far, we have assumed for simplicity that a given repressor binds different promoter sites with the same affinity (for example, $\bar{R}_1$ binds the promoters for $\bar{R}_2$ and $\bar{R}_3$ expression with the same affinity such that $K_{21} = K_{31} \equiv K_1$), and that protein expression is repressed with equal strengths $K_1 = K_2 = K_3 \equiv K$. While this symmetry allows us to observe the evolution of complex bifurcation diagrams, many biological systems are naturally perturbed away from symmetry. We now introduce biases into the relative repression strengths of a given protein, defined such that the model can still take a dimensionless form using the transformations that generated Eqns.~\ref{eq:ODEdimless_r1pre} -~\ref{eq:ODEdimless_r3pre} and Eqns.~\ref{eq:ODEdimless_r1} -~\ref{eq:ODEdimless_r3}. This allows us to determine the sensitivity of the bifurcation thresholds to various perturbations.

Returning to the single-inducer case of Section~\ref{sec:inducer1}, suppose that we keep all parameters fixed but introduce bias into regulation by the inducer-targeted $\bar{R}_1$. $\bar{R}_1$ regulates $\bar{R}_2$ and $\bar{R}_3$ expressions by binding to the respective promoter sites with affinities $K_{21}$ and $K_{31}$, respectively. We now let $K_{31}/K_{21} =\alpha$, with $K_{21}$ and all remaining dissociation constants equal to $K$ as before. The dynamics are then defined by the differential equations
\begin{align}
    \frac{d\bar{R}_{1}}{d\bar{t}} &= \frac{\bar{a}}{(1+\bar{R}_{2}^n)(1+\bar{R}_{3}^n)}-\bar{R}_{1},\label{eq:perturbAR1}\\
    \frac{d\bar{R}_{2}}{d\bar{t}} &= \frac{\bar{a}}{[1+(p_{\text{act}}(c)\bar{R}_{1})^n](1+\bar{R}_{3}^n)}-\bar{R}_{2},\label{eq:perturbAR2}\\
    \frac{d\bar{R}_{3}}{d\bar{t}} &= \frac{\bar{a}}{\Big[1+\Big(\frac{p_{\text{act}}(c)}{\alpha}\bar{R}_{1}\Big)^n\Big](1+\bar{R}_{2}^n)}-\bar{R}_{3}.  \label{eq:perturbAR3}  
\end{align}

In Section ~\ref{sec:inducer1}, we observed three distinct dynamic regimes, with decreasing complexity (i.e., number of fixed points) as the inducer concentration rose. Fig.~\ref{fig:alphathreshchange}(A) plots these shifts in dynamic regimes explicitly as changes in the number of fixed points, with the purple threshold denoting the bifurcation between distinct tristable regimes, and the green threshold denoting the bifurcation between tristable and bistable dynamics. The system sets equal interaction strengths among all repression-binding event types, i.e., $\alpha = 1$.

We now demonstrate how these bifurcations shift as we move away from symmetry. Fig.~\ref{fig:alphathreshchange}(B) reveals a narrow window of perturbation away from symmetry for which a seven-fixed-point regime exists. Expressed differently, the dark purple curve indicates a window of only $0.946 \lesssim \alpha \lesssim 1.028$ within which a system with sufficiently high concentrations of all three repressors is equally likely to stabilize to any of the three possible stable states. The dark green curve in Fig.~\ref{fig:alphathreshchange}(C) shows that the threshold separating tristable from bistable dynamics is little affected by perturbations in $\alpha \leq 1$, where $\bar{R}_1$ has a weaker affinity for the $\bar{R}_2$ promoter than for the $\bar{R}_3$ promoter. Favoring affinity to $\bar{R}_2$ for $\alpha > 1$, however, begins to affect the threshold inducer concentration noticeably. At $\alpha \gtrsim 1.42$, the system becomes bistable regardless of inducer concentration.

Combining the information in dark purple and dark green in Fig.~\ref{fig:alphathreshchange}(B) and (C), we conclude that for $\alpha < 0.95$ and $1.028\lesssim \alpha \lesssim 1.42$ only two dynamic regimes are accessible (with the seven-fixed point regime no longer viable at any inducer concentration). The perturbation threshold beyond which tristability of any kind is no longer possible occurs at $\alpha \approx 1.42$.

\begin{figure*}
    \centering
    \includegraphics[width=0.9\textwidth]{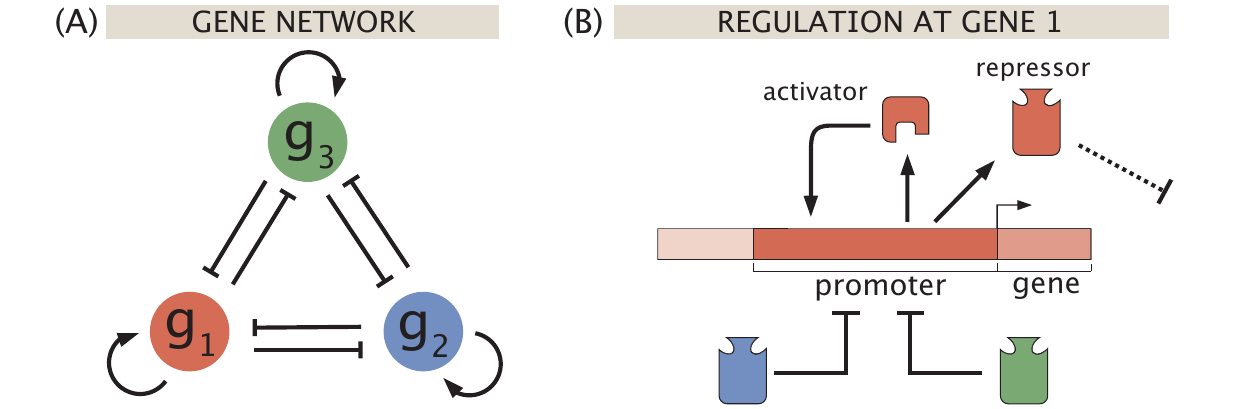}
    \caption{Gene expression in the three-gene toggle switch with self-activation. (A) Network of three mutually-repressing genes $g_i$, each producing a protein at average concentration $R_i$ that can either repress expression of $R_{j\neq i}$ or activate its own expression. (B) Regulatory binding and expression at the promoter site for gene $1$. Comparable illustrations would apply for genes $2$ and $3$.}
    \label{fig:3geneselfactdiagram}
\end{figure*}

Alternatively, we can consider a perturbation bias that affects regulation by a repressor that is not targeted by an inducer, such as $\bar{R}_2$. In this case, setting $K_{32}/K_{12} = \alpha$ and all remaining coefficients (including $K_{12}$) to $K$, we obtain the equations
\begin{align}
    \frac{d\bar{R}_{1}}{d\bar{t}} &= \frac{\bar{a}}{(1+\bar{R}_{2}^n)(1+\bar{R}_{3}^n)}-\bar{R}_{1},\\
    \frac{d\bar{R}_{2}}{d\bar{t}} &= \frac{\bar{a}}{[1+(p_{A}(c)\bar{R}_{1})^n](1+\bar{R}_{3}^n)}-\bar{R}_{2},\\
    \frac{d\bar{R}_{3}}{d\bar{t}} &= \frac{\bar{a}}{[1+(p_{A}(c)\bar{R}_{1})^n]\Big[1+\Big(\frac{\bar{R}_{2}}{\alpha}\Big)^n\Big]}-\bar{R}_{3}.
\end{align}

The light purple and green curves in Fig.~\ref{fig:alphathreshchange}(B) and (C) indicate the change in thresholds as $\alpha$ rises and alters the regulatory bias of $\bar{R}_2$. Fig.~\ref{fig:alphathreshchange}(B) shows that the change in this threshold as $\alpha$ increases from $1$, such that $\bar{R}_2$ has a stronger binding affinity for the promoter of the induced $\bar{R}_1$ than for $\bar{R}_3$, is comparable to the change observed in dark purple as described by Eqns.~\ref{eq:perturbAR1} -~\ref{eq:perturbAR3}. Decreasing $\alpha$ to favor stronger binding to the promoter for the non-induced $\bar{R}_3$, however, dampens the system's response to bias. The system denoted in light purple thus has a larger window of perturbation away from symmetry where the seven-fixed-point tristable regime is possible, with the window now favoring regulation of the non-induced over the induced repressor's expression.

The light green curve in Fig.~\ref{fig:alphathreshchange}(C), on the other hand, indicates that perturbation bias aimed to affect regulation by the non-induced repressor $\bar{R}_2$ in fact has no effect on the threshold between tristability and bistability until $\alpha \gtrsim 1.45$. We conclude that, unlike the first instance targeting the induced repressor, perturbations targeting a non-induced repressor generally influence only how complex the system's tristable dynamics can become.

\section{The three-gene toggle switch with self-activation}

The systems analyzed thus far focus exclusively on mutual repression. Toggle switches, however, can also be modeled with gene products not only repressing other genes but also stimulating their own expressions through self-activation, as shown in Fig.~\ref{fig:3geneselfactdiagram}(A). Fig.~\ref{fig:3geneselfactdiagram}(B) highlights protein production and regulatory binding at a given promoter site (in this case for gene $g_1$), where the protein produced can act either as a repressor targeting other genes or as an activator of its own expression. These dual repression and self-activation capabilities are observed across a range of natural gene network motifs with varying complexity, including applications in the bacteriophage lambda switch~\cite{JohnsonPtashne:1981,Ptashne2004,Shea1985}, stem cell and developmental differentiation~\cite{HuangEnver:2007}, and mammalian cell cycle progression~\cite{Yao2011}. We now consider the effect of self-activation on stability in the three-gene context.

The following sections assume, as in our previous baseline model from Section~\ref{subsec:baseline}, that different repressors can bind non-exclusively at a given gene promoter site to regulate expression. There are two possible ways, however, to incorporate activator binding. First, we analyze cases in which activators compete with repressors for the promoter such that an activator cannot bind if any repressor is bound, and vice versa. Section~\ref{sec:compB} then explores non-exclusive binding, where both an activator and a repressor can bind simultaneously at a given promoter site such that repressors moderate the effect of activation.

\begin{figure*}
    \centering
    \includegraphics[width=0.75\textwidth]{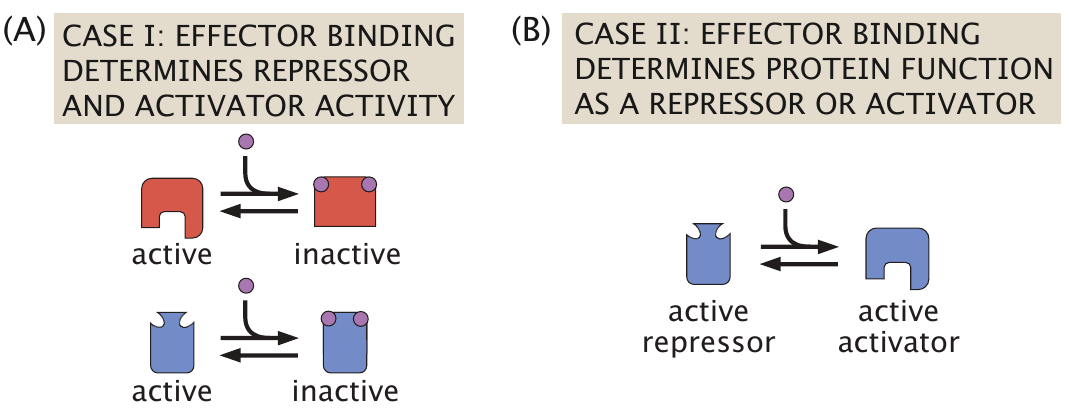}
    \caption{Two approaches to defining proteins as repressing or activating, depending on the role of effector binding. (A) Effector binding renders proteins inactive regardless of function because the protein's binding site to the DNA independently confers repressor vs. activator identity. (B) Effector binding directly determines whether a protein exists in a repressing or activating configuration, with binding transforming an active repressor to an active activator.}
    \label{fig:3geneselfact_effector}
\end{figure*}

Given that all proteins in this self-activating network can act as either repressors or activators, we are interested not only in whether they bind competitively or non-exclusively to the DNA, but also in determining what defines a protein's activity as a repressor or an activator. To do this, we highlight two biological mechanisms, and how protein activities in each case are tuned by distinct quantitative roles for effector binding. In Fig.~\ref{fig:3geneselfact_effector}(A), it is the binding site identity alone that determines a protein's function. For example, if a protein binds to its own gene's promoter site then it functions as an activator, and otherwise as a repressor. Activator proteins are then defined by the same probabilistic effector binding that captures repressor activity, with both protein types rendered inactive by effector binding. We refer to this for the remainder of our study as Case I. 

On the other hand, Fig.~\ref{fig:3geneselfact_effector}(B) depicts a setting in which effector binding directly determines protein function. Since effector binding induces a protein conformational change (favoring expression in the inducer case), we propose in this case that effector binding alters a protein's configuration from that of an active repressor to that of an active activator. We refer to this for the remainder of our study as Case II. Past precedent exists for modeling such effector-driven dual function transcription factors. For the arabinose operon, for example, the presence of arabinose induces a conformational change in AraC and thus determines its interaction with the DNA as either a repressor or an activator~\cite{Englesberg1965,Lee1987,Schleif2010}. Effector molecules can also indirectly control transcription factor function by determining whether transcription factors recruit corepressors or coactivators when bound to the DNA, such as for steroid hormone receptors~\cite{Shiau1998,Shang2000}.

We demonstrate in the following discussion how these interpretations of effector-regulated activity lead to distinct dynamics for competitive repressor-activator binding and in particular for non-exclusive binding. In comparing the two effector mechanisms, we continue to define thermodynamic states and weights that do not explicitly involve long-range binding through DNA looping. Models incorporating such looping have been considered for simpler single operons through grand canonical ensemble interpretations ~\cite{vilarSaiz2013,landman2017self}, and could be adapted to our structure. Our discussion here, however, will continue to focus on our three gene systems, but now with self-activation.

\subsection{Competitive repressor-activator binding}
\begin{figure*}
    \centering
    \includegraphics[width=0.9\textwidth]{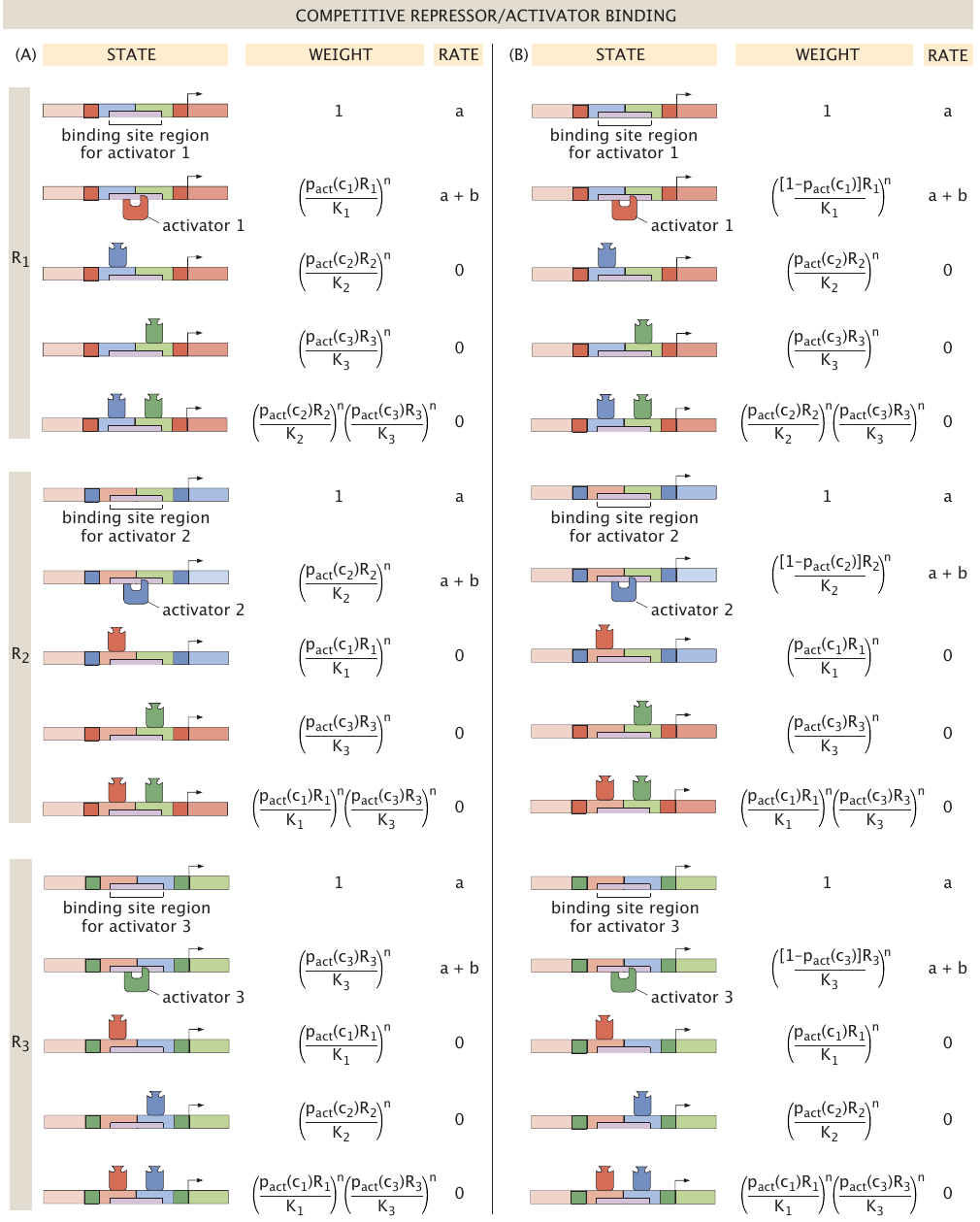}
    \caption{Expression of each protein $R_i$ in the three-gene toggle switch with self-activation and competitive binding for repressors and activators. (A) Competitive binding with the role of effectors defined as in Fig.~\ref{fig:3geneselfact_effector}(A) (Case I). (B) Competitive binding with the role of effectors defined as in Fig.~\ref{fig:3geneselfact_effector}(B) (Case II).}
    \label{fig:statesweights_comp}
\end{figure*}

We begin by considering competitive binding, with thermodynamic states, weights, and rates as depicted in Fig.~\ref{fig:statesweights_comp}. If we first assume that the binding site determines a protein's role as a repressor or an activator, then the activity of a given protein $R_i$ tuned by inducer concentration $c_i$ depends on the probability $p_{\text{act}}(c_i)$, as shown in Fig.~\ref{fig:statesweights_comp}(A). In line with the baseline model, a given protein is expressed at a maximal rate $a$ in the absence of bound regulating proteins. A bound activator, however, now increases expression to $a + b$. From these states and weights, we then define the dynamics of protein expression by
\begin{small}
\begin{align}
    \frac{dR_{1}}{dt} &= \frac{a + (a+b)(\frac{R_{1}}{K_{1}})^n}{1+(\frac{R_{1}}{K_{1}})^n + (\frac{R_{2}}{K_{2}})^n + (\frac{R_{3}}{K_{3}})^n + (\frac{R_{2}}{K_{2}})^n(\frac{R_{3}}{K_{3}})^n}-\frac{R_1}{\tau},\\
    \frac{dR_{2}}{dt} &= \frac{a+(a+b)(\frac{R_{2}}{K_{2}})^n}{1+(\frac{R_{1}}{K_{1}})^n + (\frac{R_{2}}{K_{2}})^n + (\frac{R_{3}}{K_{3}})^n + (\frac{R_{2}}{K_{2}})^n(\frac{R_{3}}{K_{3}})^n}-\frac{R_2}{\tau},\\
    \frac{dR_{3}}{dt} &= \frac{a + (a+b)(\frac{R_{3}}{K_{3}})^n}{1+(\frac{R_{1}}{K_{1}})^n + (\frac{R_{2}}{K_{2}})^n + (\frac{R_{3}}{K_{3}})^n + (\frac{R_{2}}{K_{2}})^n(\frac{R_{3}}{K_{3}})^n}-\frac{R_3}{\tau}.
\end{align}
\end{small}
By transforming $\bar{R}_i = R_i/K_1$, $\bar{t} = t/\tau$, $\bar{a} = \tau a/K_1$, and $\bar{b} = \tau b/K_1$, we obtain the dimensionless form
\begin{small}
\begin{align}
    \frac{d\bar{R}_1}{d\bar{t}} &= \frac{\bar{a} + (\bar{a} + \bar{b})\bar{R}_1^n}{1+\bar{R}_1^n + (\frac{\bar{R}_2}{K^{(2)}})^n + (\frac{\bar{R}_3}{K^{(3)}})^n + (\frac{\bar{R}_2}{K^{(2)}})^n(\frac{\bar{R}_3}{K^{(3)}})^n} - \bar{R}_1, \label{eq:Model1R1}\\
    \frac{d\bar{R}_2}{d\bar{t}} &= \frac{\bar{a} + (\bar{a} + \bar{b})(\frac{\bar{R}_2}{K^{(2)}})^n}{1+\bar{R}_1^n + (\frac{\bar{R}_2}{K^{(2)}})^n + (\frac{\bar{R}_3}{K^{(3)}})^n + (\frac{\bar{R}_2}{K^{(2)}})^n(\frac{\bar{R}_3}{K^{(3)}})^n} - \bar{R}_2, \label{eq:Model1R2}\\
    \frac{d\bar{R}_3}{d\bar{t}} &= \frac{\bar{a} + (\bar{a} + \bar{b})(\frac{\bar{R}_3}{K^{(3)}})^n}{1+\bar{R}_1^n + (\frac{\bar{R}_2}{K^{(2)}})^n + (\frac{\bar{R}_3}{K^{(3)}})^n + (\frac{\bar{R}_2}{K^{(2)}})^n(\frac{\bar{R}_3}{K^{(3)}})^n} - \bar{R}_3,\label{eq:Model1R3}
\end{align}
\end{small}
\par\noindent where $K^{(2)} = K_2/K_1$ and $K^{(3)} = K_3/K_1$. To simplify, we impose equal binding affinities such that $K^{(2)} = K^{(3)} = 1$, and examine how the system's dynamic landscape evolves with increasing inducer concentration and rising activating bias in expression.

\begin{figure*}
    \centering
    \includegraphics[width=\textwidth]{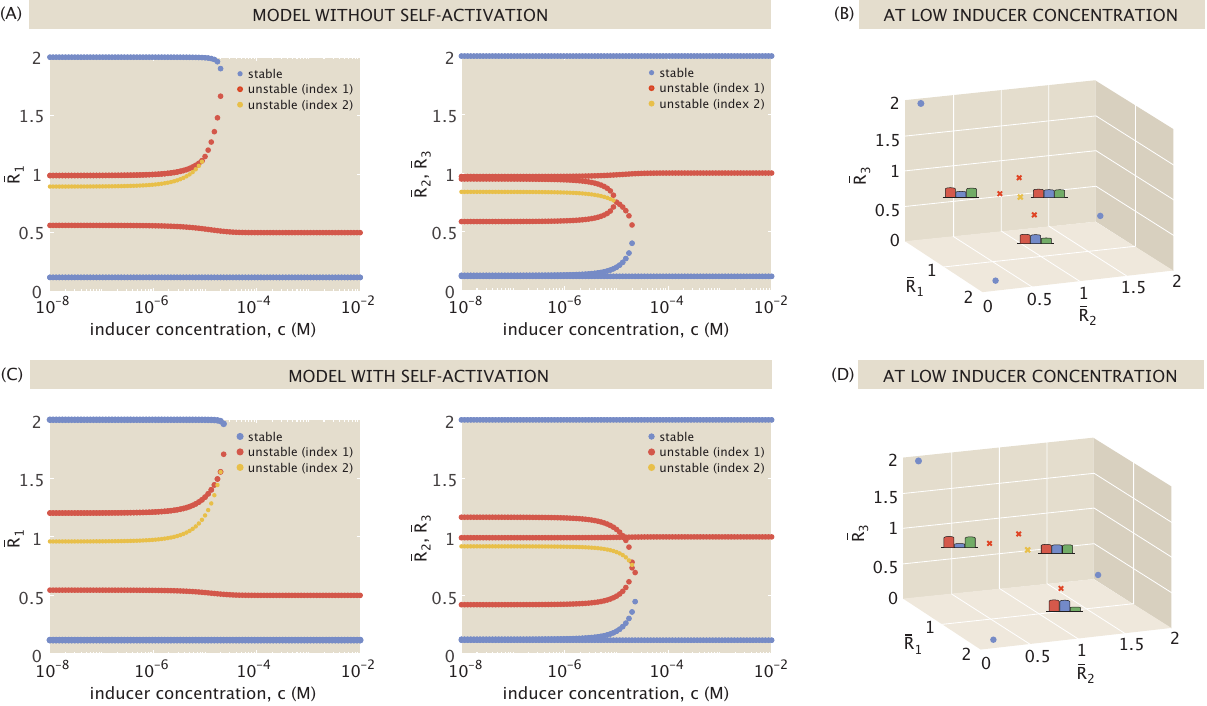}
    \caption{Comparison of the dynamics for the three-gene toggle switch without and with the presence of self-activation. In both settings, the Hill coefficient is $n = 4$ and we set the same maximal production level. (A) Bifurcation diagrams for $\bar{R}_1$, $\bar{R}_2$, and $\bar{R}_3$ steady-state expressions without self-activation as a function of inducer concentration, as previously shown in Fig.~\ref{fig:inducer1}(A). Maximal production is set at $\bar{a} = 2$. (B) Fixed points at a low inducer concentration, with labels highlighting the index-2 saddle and the induced $\bar{R}_1$-dominant index-1 saddles. (C) Bifurcation diagrams for steady-state expression with self-activation as defined by Eqns.~\ref{eq:Model1R1} -~\ref{eq:Model1R3}. Rates are set at $\bar{a} = 2$ and $\bar{b} = 0$ for maximal production satisfying $\bar{a} + \bar{b} = 2$. (D) Fixed points at a low inducer concentration.}
    \label{fig:OGvModel1}
\end{figure*}

\begin{figure*}
    \centering
    \includegraphics[width=\textwidth]{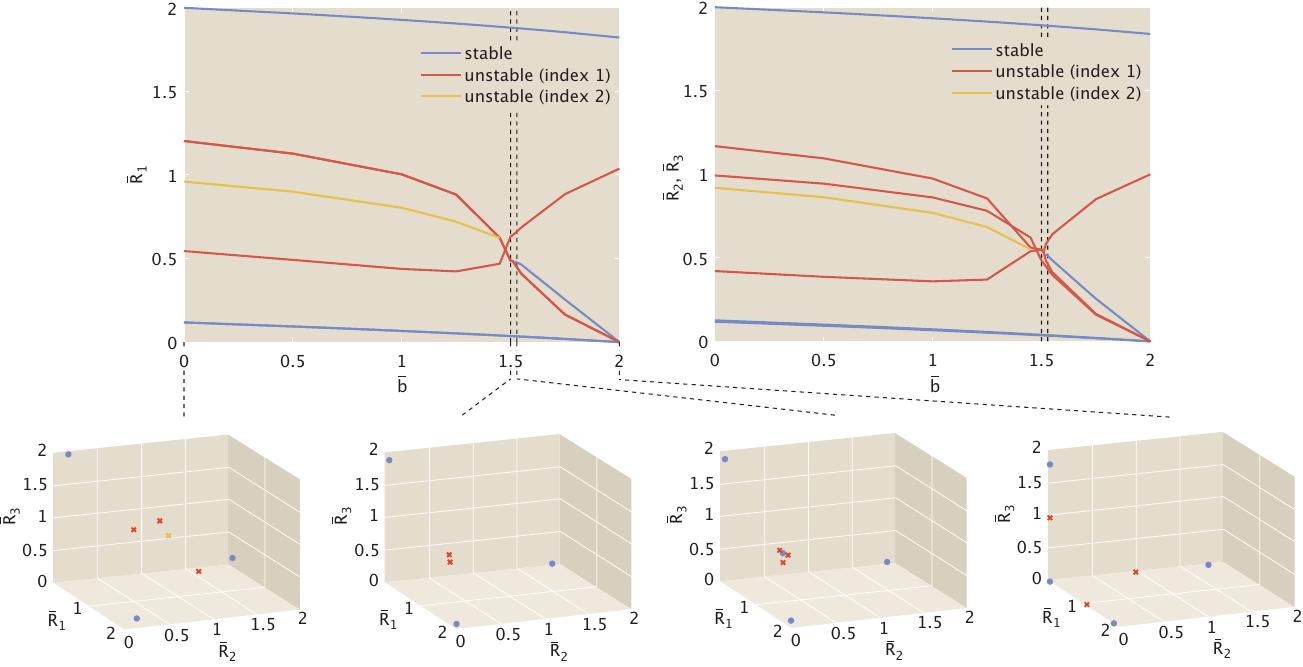}
    \caption{Bifurcation diagrams for $\bar{R}_1$, $\bar{R}_2$, and $\bar{R}_3$ expression at a low inducer concentration and at fixed $\bar{a} + \bar{b} = 2$ as a function of increasing activation strength ($\bar{b}$). The corresponding sets of fixed points are shown in three-dimensional expression space beneath, specifically highlighting each distinct dynamic phase for $(\bar{a},\bar{b}) = (2,0)$, $(\bar{a},\bar{b}) = (0.5, 1.5)$, $(\bar{a},\bar{b}) = (0.48, 1.52)$, and $(\bar{a},\bar{b}) = (0, 2)$.}
    \label{fig:Model1_b}
\end{figure*}

\subsubsection{Induction in Case I}\label{sec:comp}

We introduce self-activation in the induced setting of Case I from Fig.~\ref{fig:statesweights_comp}(A) to observe how it alters the dynamics of the baseline model. Given the states and weights listed and a single inducer $c$ targeting $\bar{R}_1$, Eqns.~\ref{eq:Model1R1} -~\ref{eq:Model1R3} become
\begin{align}
    \frac{d\bar{R}_1}{d\bar{t}} &= \frac{\bar{a} + (\bar{a} + \bar{b})[p_{\text{act}}(c)\bar{R}_1]^n}{1+[p_{\text{act}}(c)\bar{R}_1]^n + \bar{R}_2^n + \bar{R}_3^n + \bar{R}_2^n\bar{R}_3^n} - \bar{R}_1, \label{eq:Model1R1A}\\
    \frac{d\bar{R}_2}{d\bar{t}} &= \frac{\bar{a} + (\bar{a} + \bar{b})\bar{R}_2^n}{1+[p_{\text{act}}(c)\bar{R}_1]^n + \bar{R}_2^n + \bar{R}_3^n + \bar{R}_2^n\bar{R}_3^n} - \bar{R}_2, \label{eq:Model1R2A}\\
    \frac{d\bar{R}_3}{d\bar{t}} &= \frac{\bar{a} + (\bar{a} + \bar{b})\bar{R}_3^n}{1+[p_{\text{act}}(c)\bar{R}_1]^n + \bar{R}_2^n + \bar{R}_3^n + \bar{R}_2^n\bar{R}_3^n} - \bar{R}_3.\label{eq:Model1R3A}
\end{align}
To allow direct comparison with the original single-inducer model in Eqns.~\ref{eq:inducer1_r1} -~\ref{eq:inducer1_r3}, we again set $n = 4$ and now choose maximal expression to satisfy $\bar{a} + \bar{b} = 2$.

Fig.~\ref{fig:OGvModel1} compares the dynamics that emerge. Panel (A) plots the bifurcations for the baseline setting shown earlier in Fig.~\ref{fig:inducer1}, with panel (B) highlighting the fixed points at a low inducer concentration in three-dimensional expression space. Panel (C) plots the bifurcations for the comparable system with self-activation in Eqns.~\ref{eq:Model1R1A} -~\ref{eq:Model1R3A}, with panel (D) again highlighting the low inducer regime. While any number of possible $(\bar{a},\bar{b})$ pairs can satisfy $\bar{a} + \bar{b} = 2$, for closest comparison we choose the limit case of $\bar{a} = 2$ and $\bar{b} = 0$, where weak activator binding has a negligible impact on expression.

Fig.~\ref{fig:OGvModel1}(C) shows that self-activation does not affect stable expression levels or the system's dynamics at high inducer concentrations. It does, however, influence the remaining saddle points, further accentuating expression of the dominant proteins. For example, the index-2 saddle becomes more centered in expression space at the half-maximal expression level. Also, for each index-1 saddle that favors $\bar{R}_1$, the dominant protein expression levels become more pronounced compared to those that are suppressed. This means that each saddle point allowing significant $\bar{R}_1$ expression is further from the stable points it dynamically links. If the system were to begin with half-maximal expression of each protein, for example, any trajectories leaning toward the $\bar{R}_1-\bar{R}_2$ or $\bar{R}_1-\bar{R}_3$ switches would take longer to stabilize because they must travel further to approach an index-1 saddle before veering off toward a stable point. We can think of these stable states as lying in deeper potential wells, making the related systems less likely to transition from one stable state to another.

\begin{figure*}
    \centering
    \includegraphics[width=0.95\textwidth]{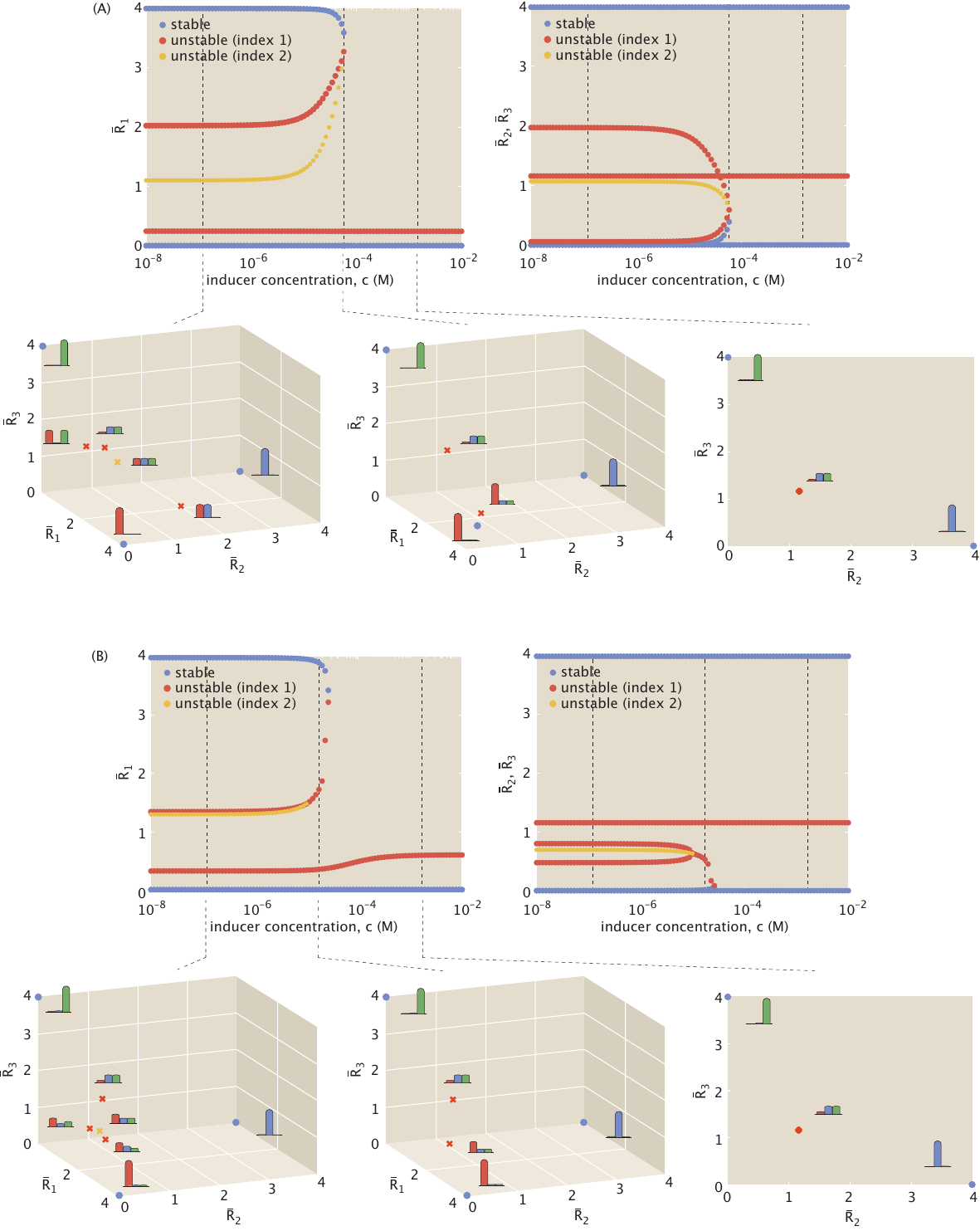}
    \caption{Induction Cases I and II for competitive repressor-activator binding, with $\bar{a} = 2$, $\bar{b} = 2$. Each panel plots the bifurcation diagrams for $\bar{R}_1$, $\bar{R}_2$, and $\bar{R}_3$ steady-state expressions as a function of inducer concentration. The fixed points are visualized in expression space beneath the corresponding bifurcations at low, intermediate, and high inducer concentrations. The low and high inducer concentrations chosen are the same in (A) and (B), and the intermediate concentrations selected highlight the same intermediate dynamic regime. (A) The competitive repressor-activator binding case in which inducer binding controls transcription factor activity regardless of its ultimate function as a repressor or activator (Case I). (B) The competitive repressor-activator binding case in which inducer binding transforms a protein from a repressing conformation to an activating form (Case II).}
    \label{fig:Model1vs3_b2}
\end{figure*}

As one might expect, the differences between the baseline model and our parametrization of the self-activation model are subtle, but the key advantage of self-activation lies in the additional tunability that becomes possible through the additive expression effect from $\bar{b}$. Increasing the strength of activation while maintaining a constant maximal expression level $\bar{a} + \bar{b}$ can lead to significant dynamic transformations. Fig.~\ref{fig:Model1_b} fixes the system at a low inducer concentration for $\bar{a} + \bar{b} = 2$, and plots the bifurcation diagram with dynamics tuned as a function of activation $\bar{b}$. As $\bar{b}$ increases from zero, all fixed points shift toward lower expression while retaining the typical dynamic profiles observed in Fig.~\ref{fig:OGvModel1}. Plotting how the fixed points shown for $(\bar{a},\bar{b}) = (2,0)$ would change with increasing inducer concentration, for example, would generate two dynamic phase shifts to arrive at a bistable switch between $\bar{R}_2$ and $\bar{R}_3$ expression, with results comparable to those shown in Fig.~\ref{fig:inducer1}.

With sufficiently strong activation at $\bar{b}\approx 1.47$, however, the system no longer retains the index-2 saddle. This is evidenced by the fixed point plot highlighted at $\bar{b} = 1.5$ in Fig.~\ref{fig:Model1_b}. Moreover, $\bar{b} = 1.52$ marks a threshold from tristable to \emph{quadristable} expression, where the index-2 fixed point transforms into a stable point. Rather than being comprised of a set of switches between $\bar{R}_i$ and $\bar{R}_j$ expression, the high $\bar{b}$ regime now brings together a set of switches between ``on'' and ``off'' expression for each protein. Note that, as the bifurcation point, $\bar{b} = 1.52$ shares properties between the two dynamic phases it straddles. As shown in Fig.~\ref{fig:Model1_b}, this point allows the quadristable dynamics observed for higher $\bar{b}$ at low inducer concentration, while still reducing to a single bistable switch at high inducer concentration (as is characteristic of the small $\bar{b}$ regime). Systems approaching the limit where expression can only occur in the presence of bound activator, on the other hand, as in the highlighted example $(\bar{a},\bar{b}) = (0,2)$ of Fig.~\ref{fig:Model1_b}, will reduce down to tristability at high inducer concentration, allowing either $\bar{R}_2$-dominant expression, $\bar{R}_3$-dominant expression, or none.

\subsubsection{Induction in Case II}\label{sec:compB}

We now compare induction Case I above to the dynamics arising in an alternative setting, Case II, where the effector molecule binding determines whether a protein is active as a repressor or an activator. Fig.~\ref{fig:statesweights_comp}(B) outlines the thermodynamic states and weights in this setting, from which we model the system with the differential equations
\begin{align}
    \frac{d\bar{R}_1}{d\bar{t}} &= \frac{\bar{a} + (\bar{a} + \bar{b})[(1-p_{\text{act}}(c))\bar{R}_1]^n}{1+[(1-p_{\text{act}}(c))\bar{R}_1]^n + \bar{R}_2^n + \bar{R}_3^n + \bar{R}_2^n\bar{R}_3^n} - \bar{R}_1, \label{eq:Model3R1}\\
    \frac{d\bar{R}_2}{d\bar{t}} &= \frac{\bar{a} + (\bar{a} + \bar{b})\bar{R}_2^n}{1+[p_{\text{act}}(c)\bar{R}_1]^n + \bar{R}_2^n + \bar{R}_3^n + \bar{R}_2^n\bar{R}_3^n} - \bar{R}_2, \label{eq:Model3R2}\\
    \frac{d\bar{R}_3}{d\bar{t}} &= \frac{\bar{a} + (\bar{a} + \bar{b})\bar{R}_3^n}{1+[p_{\text{act}}(c)\bar{R}_1]^n + \bar{R}_2^n + \bar{R}_3^n + \bar{R}_2^n\bar{R}_3^n} - \bar{R}_3.\label{eq:Model3R3}
\end{align}
To aid in visual comparison with induction Case I, we consider systems allowing a larger maximal expression of $\bar{a} + \bar{b} = 4$. Specifically, we will set $\bar{a} = \bar{b} = 2$ for equally strong contributions from the basal and activated expression levels.

Fig.~\ref{fig:Model1vs3_b2} compares the system defined by induction Case I, now for $\bar{a} = \bar{b} = 2$, with the corresponding model in panel (B) for the alternative induction approach highlighted in Eqns.~\ref{eq:Model3R1} -~\ref{eq:Model3R3}. We observe that in the low inducer concentration regime, while the dynamical structure remains the same, the saddle points allowing $\bar{R}_1$ expression in panel (B) are now skewed toward favoring $\bar{R}_1$ expression. In this regime, a low inducer concentration implies a high probability $p_{\text{act}}(c)$ of $\bar{R}_1$ acting as a repressor, and a low probability $1-p_{\text{act}}(c)$ of acting as an activator. With $[1-p_{\text{act}}(c)]\bar{R}_1 \:\rightarrow\: 0$ in Eqn.~\ref{eq:Model3R1}, $\bar{R}_1$ expression is limited only by regulatory repression from $\bar{R}_2$ and $\bar{R}_3$, while the remaining two repressors are additionally regulated by $p_{\text{act}}(c)\bar{R}_1$. The resulting skewness of the central fixed point cluster to favor $\bar{R}_1$ places the $\bar{R}_2$ and $\bar{R}_3$-dominant stable states in deeper potential wells. It thus requires a less significant perturbation to transition out of the $\bar{R}_1$-dominant stable state into one of the other stable states.

Since all $\bar{R}_1$-expressing saddle points are spatially closer together in the lower panel compared to the upper panel of Fig.~\ref{fig:Model1vs3_b2}, this indicates that Case II of induction does not require as high an inducer concentration to transition to an intermediate dynamic regime. Finally, at a sufficiently high inducer concentration, the saddle of the resulting bistable switch shifts the saddle point toward a slightly higher $\bar{R}_1$ concentration, which serves to deepen the potential well between the $\bar{R}_2$ and $\bar{R}_3$-dominant stable states compared to Case I.

\subsection{Non-exclusive repressor-activator binding} \label{sec:nonexc}

\begin{figure*}
    \centering
    \includegraphics[width=\linewidth]{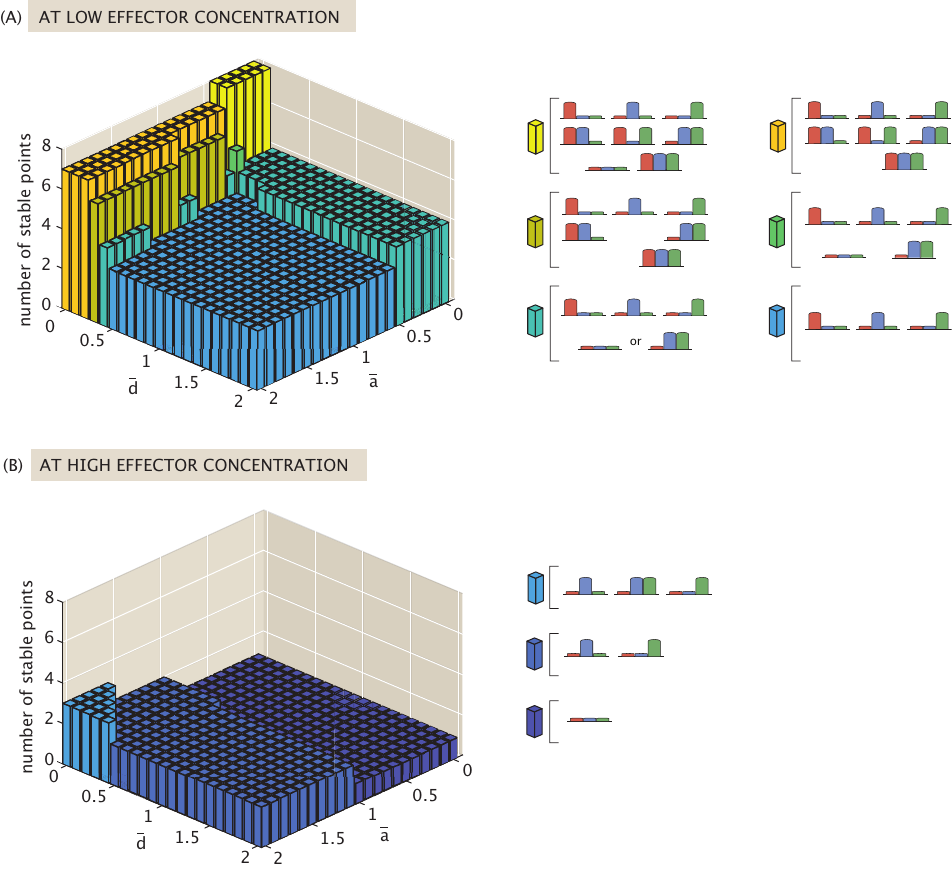}
    \caption{Number of stable fixed points in induction Case I of non-exclusive repressor-activator binding. The plots highlight stable points at low ($10^{-8}$ M) and high ($10^{-2}$ M) effector concentrations in panels (A) and (B), respectively, for varying rate parameters $\bar{a}$, $\bar{b}$, and $\bar{d}$, where $\bar{a} + \bar{b} = 2$ and $\bar{d} \leq \bar{a}+\bar{b}$. The diagrams on the right hand side of panel (A) represent the types of stable points observed in each regime at a low effector concentration, i.e., with eight (yellow), seven (orange), six (brown), five (green), four (teal), or three (blue) stable points. The diagrams on the right side of panel (B) represent the types of stable points observed in each regime at a high effector concentration, i.e., with three (blue), two (dark blue), or one (purple) stable point(s).}
    \label{fig:nonexcA_fpstartend}
\end{figure*}

We now turn to systems in which repressors and activators bind non-exclusively to a gene's regulatory region. This means that at a given promoter site, a gene's expression can be regulated not only by the presence of one or both repressors produced by the remaining two genes, but also by the possible additional presence of bound activator. As in the competitive binding case shown in Fig.~\ref{fig:statesweights_comp}, this system with non-exclusive binding retains the same possible states for a given promoter site, namely (i) the state with no bound transcription factors (with an expression rate $a$), (ii) states with one or both possible repressors bound (with an expression rate $0$), and (iii) states with activator bound alone (with an expression rate $a + b$). Additional states account for activated expression in the presence of one or both available repressors, and in these states, activated expression is suppressed by the presence of bound repressor defined by a corresponding rate $a + b - d$.

Upon deriving a set of differential equations in dimensionless form for the expression of proteins $\bar{R}_1$, $\bar{R}_2$, and $\bar{R}_3$ from the corresponding thermodynamic states and weights (see Appendix~\ref{app:nonexc} for details), we have, as in the competitive binding setting, a system defined by basal expression $\bar{a}$ and additional activated expression $\bar{b}$, with the maximal expression level constrained once again to $\bar{a} + \bar{b} = 2$ to facilitate comparison across models. Given the non-exclusive binding conditions modeled, expression now also depends on the strength of repression from $\bar{d}$, where $\bar{d}\leq \bar{a} + \bar{b}$.

The following subsections focus on the impact of effector concentration and the relative strengths of activation and repression on dynamic stability. This allows us to determine the range of dynamic landscapes possible within the parameter space under different interpretations for effector activity, and the sensitivity of the resulting dynamics and bifurcations to each of the rate parameters.

\subsubsection{Induction in Case I}\label{sec:nonexcA}

We first consider Case I from Fig.~\ref{fig:3geneselfact_effector}(A), in which the binding of the effector molecule renders proteins inactive regardless of whether they are functioning as repressors or activators. Continuing our convention of choosing $\bar{R}_1$ as the target of effector binding, the concentration of $\bar{R}_1$ is scaled by the probability of activity $p_{\rm act}(c)$, regardless of function, where we assume that all proteins interact with effectors through the same MWC model and binding affinities. Fig.~\ref{fig:statesweights_nonexcA} and Eqns.~\ref{eq:barR1nonexcA} -~\ref{eq:barR3nonexcA} of Appendix~\ref{app:nonexc} define the corresponding states and weights along with the resulting dimensionless differential equations incorporating the role of $p_{\rm act}(c)$.

\begin{figure}
    \centering
    \includegraphics[width=\columnwidth]{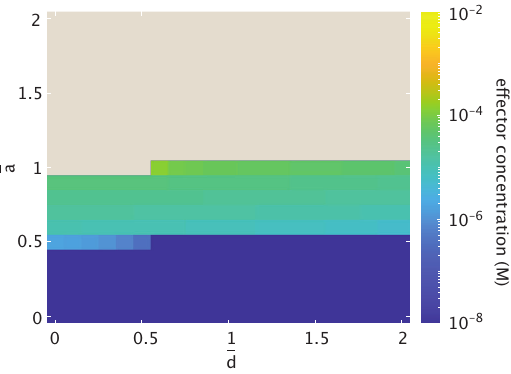}
    \caption{Heatmap tracking the minimum effector concentration (M) at which the stable state with no expression can exist with non-exclusive activator-repressor binding (induction Case I). Measurements span the range of rate parameters allowed by the constraints $\bar{a} + \bar{b} = 2$ and $\bar{d} \leq \bar{a} + \bar{b}$, with the x-axis denoting increasing strength of repression $\bar{d}$, and the y-axis denoting decreasing strength of activation with increasing $\bar{a}$. The beige region represents the parameter space in which a stable state without expression does not exist at any effector concentration.}
    \label{fig:nonexcA_stabstart000}
\end{figure}

\begin{figure*}
    \centering
    \includegraphics[width=0.9\textwidth]{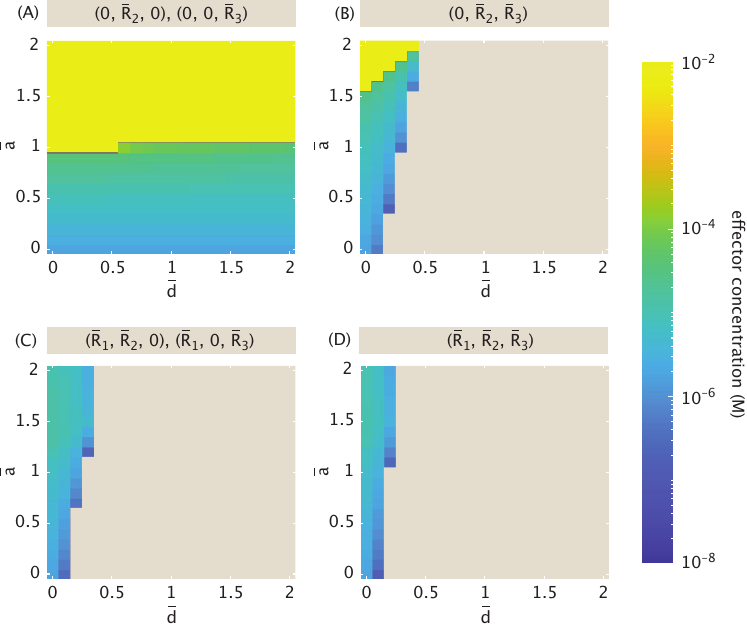}
    \caption{Heatmaps tracking the maximum effector concentrations ($M)$ (bifurcation thresholds) at which stable states can exist for the non-exclusive activator-repressor binding model described in Fig.~\ref{fig:3geneselfact_effector}(A) (Case I). The heatmaps track (A) stable states expressing one of the genes that does not directly interact with an effector, i.e., $(0,\bar{R}_2,0)$ and $(0,0,\bar{R}_3)$, (B) the stable state in which the two genes not impacted by effector are expressed, i.e., $(0,\bar{R}_2,\bar{R}_3)$, (C) stable states expressing two genes including the effector-controlled $\bar{R}_1$, and (D) the stable state expressing all three genes. Measurements span the rate parameters allowed by the constraints $\bar{a} + \bar{b} = 2$ and $\bar{d} \leq \bar{a} + \bar{b}$, with the x-axis denoting increasing repression strength $\bar{d}$, and the y-axis denoting decreasing activation with increasing $\bar{a}$. Beige represents the region of parameter space in which the stable state cannot exist at any effector concentration, and the yellow-shaded region indicates where the stable state is possible at all effector concentrations.}
    \label{fig:nonexcA_stabend}
\end{figure*}

Evaluating the fixed points, we find that allowing non-exclusive binding expands the possible set of stable states to the maximum combinatorial number, i.e., no expression, one gene expressing, two genes expressing, or all three expressing. Determining the likelihood of each state being accessible within the allowable rate parameter space offers insights into how this more complex regulatory architecture prioritizes expression.

Fig.~\ref{fig:nonexcA_fpstartend}(A) tracks the number of possible stable states that exist at low effector concentration ($c = 10^{-8}$ M). The maximum number of stable states occurs in the yellow-shaded region where $\bar{d}\approx 0$ and $\bar{a} < 0.5$ ($\bar{b} > 1.5$). This represents systems in which repressor binding has negligible effects on expression, and where gene expression thus relies heavily on bound activators.

As $\bar{a}$ and $\bar{d}$ rise, the complexity of the dynamic landscape decreases, with the number of possible stable points falling until $\bar{a}\gtrsim 0.5$ and $\bar{d}\gtrsim 0.5$. This threshold combination is sufficient to render the system tristable at low effector concentrations, and is similar to what we observed in the previous competitive binding models of Sections~\ref{sec:comp} and \ref{sec:compB}.

Other than the three possible single-gene expression states, which exist at low effector concentrations for all possible rate parameter combinations, the stable state that survives across the broadest allowable parameter space is the one in which both $\bar{R}_2$ and $\bar{R}_3$ dominate. Such a state becomes no longer possible only if $\bar{a}\gtrsim 0.5$ and $\bar{d}\gtrsim 0.5$. The robustness of this state makes sense given that the model in question specifically tunes $\bar{R}_1$ expression via effector concentration. This targeted tuning more indirectly (and weakly) affects the existence of a stable $(0,\bar{R}_2,\bar{R}_3)$ state.  Fig.~\ref{fig:nonexcA_fpstartend}(A) also indicates that the repression parameter $\bar{d}$ drives the loss of the $(\bar{R}_1,0,\bar{R}_3)$ and $(\bar{R}_1,\bar{R}_2,0)$ states. Essentially, once repressors play an active role in regulation (no longer $\bar{d}\rightarrow 0$), dual repressor and activator activity implies that the system can only express $\bar{R}_1$ significantly in a stable state when it is the only protein being expressed.

Generally, as effector concentrations increase, the number of available stable fixed points decreases until the system reaches monostable, bistable, or tristable dynamics, as shown in Fig.~\ref{fig:nonexcA_fpstartend}(B). When $\bar{a} \lesssim 1$, such that activation contributes more strongly than basal expression, a high effector concentration suppresses all gene expression. When the system approaches the limit case of largely unregulated expression (i.e., $\bar{a} \rightarrow 2$ and $\bar{d} \rightarrow 0$) it becomes tristable, expressing either $\bar{R}_2$, $\bar{R}_3$, or both. Otherwise, in the relatively weak activation regime that remains, the system collapses to the bistable toggle switch between $\bar{R}_2$ and $\bar{R}_3$. 

Figs.~\ref{fig:nonexcA_stabstart000} and~\ref{fig:nonexcA_stabend} illustrate further how the system evolves toward these different regimes as the effector concentration rises, and how sensitive the different stable states are to these changes. The figures plot (respectively) the minimum and maximum effector concentrations that are required for different stable fixed points to exist, where Fig.~\ref{fig:nonexcA_stabstart000} specifically highlights the minimum concentration for the stable state with no expression. At the maximum concentration threshold, a bifurcation occurs when a sufficiently high effector concentration prevents the system from stabilizing to the state with no expression. The beige-shaded region denotes the part of parameter space where a given fixed point does not exist at any effector concentration, and the yellow-shaded region denotes the space where the fixed point survives at high effector concentration.

Notably, the heatmaps of Fig.~\ref{fig:nonexcA_stabstart000} and Fig.~\ref{fig:nonexcA_stabend}(A) directly overlap, indicating a close relationship between the existence of a stable state with no expression and the existence of a state that expresses one of the two proteins not impacted by the effector ($\bar{R}_2$ or $\bar{R}_3$). The existence of these three stable states is thus defined almost fully by the strength of activation as reflected by $\bar{a}$. As $\bar{a}$ rises, three regimes with distinct dynamic trends emerge upon tuning the effector concentration. When $\bar{a}\lesssim 0.5$, all three states are possible until an intermediate effector concentration is reached, beyond which only the state suppressing all expression can exist among the three. When $0\lesssim \bar{a}\lesssim 1$, we observe a tradeoff at an intermediate effector concentration, where the system's ability to stabilize to an $\bar{R}_2$ or $\bar{R}_3$-dominant state at low concentrations is swapped for the system's ability to suppress all expression. This indicates a regime of strong activation compared to basal expression, where an increase in effector concentration causes a fixed point to emerge and replace states that existed at lower concentrations. Finally, in the weak activation regime of $\bar{a}\gtrsim 1$, the system can always stabilize to an $\bar{R}_2$ or $\bar{R}_3$-dominant state at all effector concentrations, and never suppresses expression entirely. Note in all of these regimes that the strength of repression ($\bar{d}$) has negligible or no effect on the existence of the three states.

Fig.~\ref{fig:nonexcA_stabend}(B)-(D) highlights stable fixed points where more than one type of protein is expressed. The existence of such points relies on weak repression $(\bar{d} < 0.5)$. As $\bar{a}$ rises and the strength of activation decreases accordingly, we observe levels of rising effector concentrations where the stable points vanish. The heatmaps indicate that as repression strength rises, a higher value of $\bar{a}$ and thus a weaker degree of activation is needed for the system to stabilize to these states, with the state in panel (B) responding most gradually and the state in panel (D) most sharply as repression strength increases. Finally, the span of effector concentrations indicates that the bifurcation thresholds for these stable points are more responsive to changes in activation strength than those of the single-protein expression states shown in Fig.~\ref{fig:nonexcA_stabend}(A).

We conclude from Fig.~\ref{fig:nonexcA_stabend} that expressing more than one protein at a stable steady state is only possible in a system with weak repression, whereas repression has essentially no effect on expression that is not targeted by an effector ($\bar{R}_2$ and $\bar{R}_3$ here). Additionally, while weakening activation strength always expands the range of effector concentrations at which these stable fixed points can exist, only states that do not express the effector-targeted genes can exist at all effector concentrations (the yellow regions of Fig.~\ref{fig:nonexcA_stabend}). These yellow-shaded regions require weaker activation strengths compared to the strength of unregulated expression, with this requirement even stricter for the state expressing more than one protein. The results demonstrate that by tuning these parameters and testing varying effector concentrations, one can transform a system to express different numbers of genes at stable steady states depending on which combinations are most relevant for a desired function.

\subsubsection{Induction in Case II}\label{sec:nonexcB}

\begin{figure*}
    \centering
    \includegraphics[width=\textwidth]{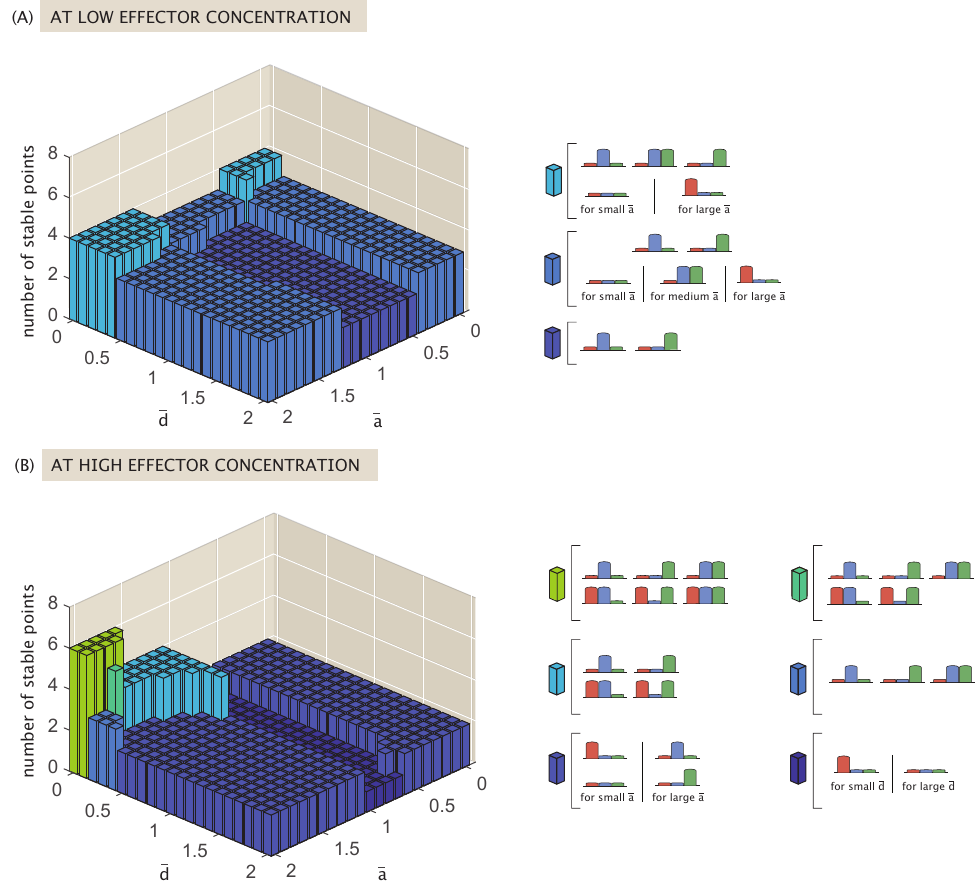}
    \caption{Number of stable fixed points in induction Case II of non-exclusive repressor-activator binding. The plots highlight stable points at low ($10^{-8}$ M) and high ($10^{-2}$ M) effector concentrations in panels (A) and (B), respectively, for varying rate parameters $\bar{a}$, $\bar{b}$, and $\bar{d}$, where $\bar{a} + \bar{b} = 2$ and $\bar{d} \leq \bar{a}+\bar{b}$. The diagrams on the right hand side of panel (A) represent the types of stable points observed in each regime at a low effector concentration, i.e., with four (blue), three (dark blue), or two (purple) stable points. The diagrams on the right side of panel (B) represent the types of stable points observed in each regime at a high effector concentration, i.e., with six (green), five (teal), four (blue), three (dark blue), two (purple), or one (dark purple) stable point(s).}
    \label{fig:nonexcB_fpstartend}
\end{figure*}

\begin{figure*}
    \centering
    \includegraphics[width=\textwidth]{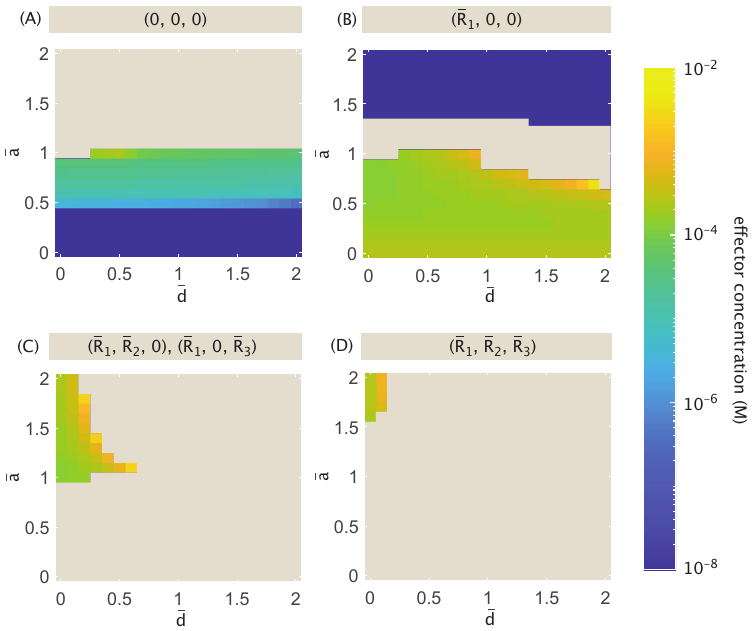}
    \caption{Heatmap tracking the minimum effector concentration (M) at which stable states can exist with the non-exclusive activator-repressor binding model described in Fig.~\ref{fig:3geneselfact_effector}(B) (Case II). Measurements span the rate parameters allowed by the constraints $\bar{a} + \bar{b} = 2$ and $\bar{d} \leq \bar{a} + \bar{b}$, with the x-axis denoting increasing repression strength $\bar{d}$, and the y-axis denoting decreasing activation with increasing $\bar{a}$. The heatmaps track (A) the stable state in which no genes are expressed, (B) the stable state in which only the effector-targeted $\bar{R}_1$ is expressed, (C) stable states expressing two genes including the effector-controlled $\bar{R}_1$, and (D) the stable state expressing all three genes. Beige represents the region of parameter space in which the stable state cannot exist at any effector concentration.}
    \label{fig:nonexcB_stabstart}
\end{figure*}

\begin{figure*}
    \centering
    \includegraphics[width=\textwidth]{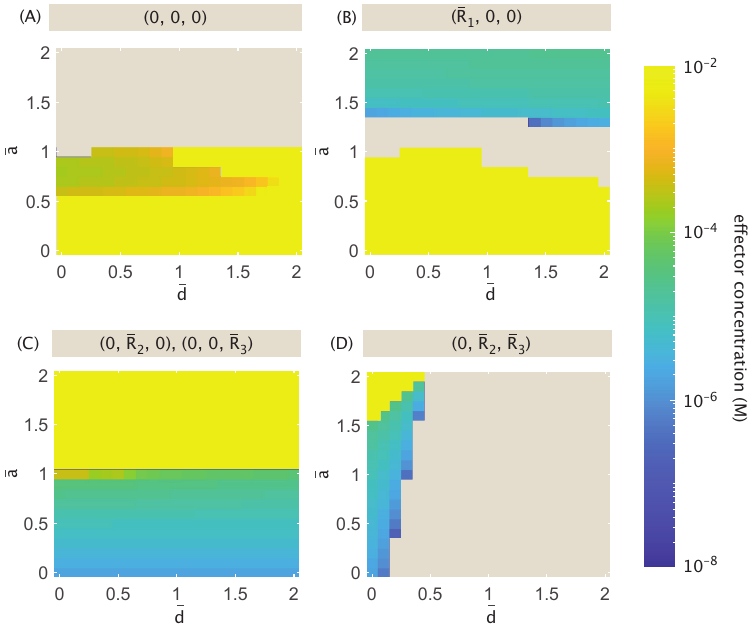}
    \caption{Heatmaps tracking the maximum effector concentrations (M) at which stable states can exist with non-exclusive activator-repressor binding model described in Fig.~\ref{fig:3geneselfact_effector}(B) (Case II). The measured values represent bifurcation thresholds for the existence of these states. The heatmaps track (A) the stable state in which no genes are expressed, (B) the stable state in which only the effector-targeted $\bar{R}_1$ is expressed, (C) stable states expressing one of the genes that does not directly interact with effector, i.e., $(0,\bar{R}_2,0)$ and $(0,0,\bar{R}_3)$, and (D) the stable state in which both genes not impacted by effector are expressed, i.e., $(0,\bar{R}_2,\bar{R}_3)$. Measurements span the rate parameters allowed by the constraints $\bar{a} + \bar{b} = 2$ and $\bar{d} \leq \bar{a} + \bar{b}$, with the x-axis denoting increasing repression strength $\bar{d}$, and the y-axis denoting decreasing activation as $\bar{a}$ rises. Beige represents the region where a stable state never exists at any effector concentration, and the yellow region indicates that the stable state is possible at all effector concentrations.}
    \label{fig:nonexcB_stabend}
\end{figure*}
An alternative interpretation for effector activity, shown in Case II of Fig.~\ref{fig:3geneselfact_effector}(B), transforms the available states and weights from the non-exclusive case such that the concentration of transcription factor $\bar{R}_1$ is scaled by $p_{\rm act}(c)$ when repressing expression of other proteins, and is scaled by $1-p_{\rm act}(c)$ when activating its own expression. (See Appendix~\ref{app:nonexc} and corresponding Fig.~\ref{fig:statesweights_nonexcB} for the thermodynamic states and weights and for the corresponding model form.)

By evaluating the resulting differential equations for fixed points, we obtain plots in Fig.~\ref{fig:nonexcB_fpstartend} highlighting the types of stable points possible at low ($10^{-8}$ M) and high ($10^{-2}$ M) effector concentrations when tuning $\bar{a}$ and $\bar{d}$, with maximal production at a fixed $\bar{a} + \bar{b}$. Fig.~\ref{fig:nonexcB_fpstartend}(A) indicates that, regardless of activation strength ($\bar{a}$) and repression ($\bar{d}$), the system can always stabilize to a state in which either $\bar{R}_2$ or $\bar{R}_3$ dominates at low effector concentrations. When $0.5 \leq \bar{a} \lesssim 1.3$ and $\bar{d} \gtrsim 0.3$ (purple-shaded region), the system can only exist as a bistable switch between these two states. When repression is weak (small $\bar{d}$), the system can also stabilize to a state expressing both genes (regardless of $\bar{a}$). Meanwhile, the strength of activation solely determines whether it is possible for the system to suppress expression altogether or to stabilize to an $\bar{R}_1$-dominant state. For sufficiently strong activation ($\bar{a} < 0.5$), it is possible for regulation to suppress all expression, and it is only under weak activation  ($\bar{a} \gtrsim 1.3$) that it is possible for the system to express the effector-targeted $\bar{R}_1$.

At a high effector concentration, Fig.~\ref{fig:nonexcB_fpstartend}(B) indicates that the most complex dynamic landscape occurs when $\bar{d}$ is very small and $\bar{a}$ is very large, i.e., when there is effectively no positive or negative regulation in the network. Such a system can stabilize to either $\bar{R}_2$ or $\bar{R}_3$-dominant states, or to a state in which any combination of two or more proteins dominates (including effector-targeted $\bar{R}_1$), as shown in the green-shaded region. Moving slightly away from this corner of parameter space, the system loses the ability to express all three genes, a change that is more sensitive to increased repression than increased activation. Increasing repression ($\bar{d}$) further while keeping $\bar{a}$ high suppresses the system's ability to express effector-targeted $\bar{R}_1$. In this case, the dynamics become comparable to a self-activated bistable switch, as in the dark blue region of Fig.~\ref{fig:nonexcB_fpstartend}(B), or the simple bistable switch between $\bar{R}_2$ and $\bar{R}_3$, as in the purple region. Increasing activation (decreasing $\bar{a}$) while keeping repression relatively weak ($\bar{d}\lesssim 0.5$), on the other hand, implies that if the system expresses more than one gene in its stable state, one of them must be the effector-targeted $\bar{R}_1$, as seen in the blue-shaded region. In a strongly activating regime with small $\bar{a}$, dynamics center on expression of $\bar{R}_1$, either stabilizing to an $\bar{R}_1$-dominant state or suppressing all expression (purple region). If $\bar{a} \rightarrow 1$ as in the dark purple region of Fig.~\ref{fig:nonexcB_fpstartend}(B), however, the system chooses one of these fates depending on $\bar{d}$, only expressing a gene if repression ($\bar{d}$) remains small.

As previously shown in Fig.~\ref{fig:nonexcA_stabstart000} and Fig.~\ref{fig:nonexcA_stabend} for induction Case I, we can now also investigate how different parametrizations for activation and repression in Case II influence the range of effector concentrations for which the different types of stable fixed points emerge. Fig.~\ref{fig:nonexcB_stabstart} tracks the minimum effector concentrations at which fixed points become possible within the parameter space, while Fig.~\ref{fig:nonexcB_stabend} tracks the maximum effector concentrations. Given that Case II differs from Case I only in its treatment of $\bar{R}_1$, it follows that Fig.~\ref{fig:nonexcA_stabend}(A)-(B) and Fig.~\ref{fig:nonexcB_stabend}(C)-(D) show no difference in the expression patterns for those stable states that exclusively express one or both of the non-targeted genes, i.e., $(0,\bar{R}_2,0)$, $(0,0,\bar{R}_3)$, and $(0,\bar{R}_2,\bar{R}_3)$. A slight difference occurs for states expressing one of the two non-targeted genes when $\bar{a} = 1$ and $\bar{d}\leq 0.5$. Under these conditions, where Case I would allow these states to exist at all effector concentrations, there is instead still an upper bound shown in Fig.~\ref{fig:nonexcB_stabend}(C) for Case II beyond which the states cannot exist. This means that when effectors determine whether a protein acts a repressor or an activator, and the system neither favors nor disfavors activation, minimizing the effect of repression is not sufficient to allow expression at all effector concentrations, and that this can only be true if the system explicitly favors activation with $\bar{a} > 1$.

States expressing the effector-targeted $\bar{R}_1$ in Case II offer a stark contrast to those in Case I. In the latter setting, the $\bar{R}_1$-dominant stable state is not included in Figs.~\ref{fig:nonexcA_stabstart000} and~\ref{fig:nonexcA_stabend} because it can exist for essentially the same range of low effector concentrations (from $10^{-8}$ M to approximately $5\times 10^{-5}$ M) regardless of $\bar{a}$ and $\bar{d}$. In Case II, however, this is only observed for $\bar{a} \gtrsim 1.4$, as seen in Fig.~\ref{fig:nonexcB_stabstart}(B) and Fig.~\ref{fig:nonexcB_stabend}(B). Further, if $\bar{a} \leq 1$ (with some exceptions), the state can only exist at larger effector concentrations ranging from approximately $10^{-4}$ M to the highest parameter value, $10^{-2}$ M. Thus, when activation is particularly strong, the phase space for $\bar{R}_1$-dominant expression spans the same range of concentrations observed for Case I. When activation is disfavored as the source of protein production compared to the basal level of expression, however, $\bar{R}_1$-dominant expression emerges as a viable stable state only at high effector concentrations lying outside of the previously established range.

Considering the stable state in which all expression is suppressed, Fig.~\ref{fig:nonexcA_stabstart000} and Fig.~\ref{fig:nonexcB_stabstart}(A) demonstrate that regardless of the effector's role in regulating $\bar{R}_1$, this type of stable state can exist only when activation is favored over the basal rate ($\bar{a}\leq 1$). The strong resemblance between these plots also indicates that, in the viable region of parameter space, tuning $\bar{a}$ and $\bar{d}$ similarly affects the minimum concentration necessary for the state to exist. We observe small differences at the two visible thresholds:  $\bar{a} = 0.5$ separates the high-activation regime in which the state always exists at low effector concentration from the regime characterized by a particular threshold effector concentration; and $\bar{a} = 1$ denotes the upper bound of $\bar{a}$ for which the $(0,0,0)$ state can exist at any effector concentration. Compared to Case I, when $\bar{a} = 1$ Case II requires an even weaker level of repression $\bar{d}$ to prevent the system from suppressing all expression.

Unlike Case I, for Case II the existence of the $(0,0,0)$ stable state does not immediately imply its viability for all effector concentrations above the minimum threshold. Fig.~\ref{fig:nonexcB_stabend}(A) shows that, in a regime of intermediate activation strength where $0.6 \leq \bar{a} \leq 1$, there is an upper bound on the effector concentration. These intermediate conditions mark a regime in which the effector concentration can be manipulated to tune stable expression off and on.

Finally, we consider stable states expressing more than one protein, including the effector-targeted $\bar{R}_1$. Whereas Fig.~\ref{fig:nonexcA_stabend}(C)-(D) indicates that these states can exist for Case I at low effector concentrations up to values between $10^{-6}$ M and $10^{-5}$ M for low repression $\bar{d}$, Fig.~\ref{fig:nonexcB_stabstart}(C)-(D) indicates an overlapping regime of existence but with distinctly different dependencies on both rate parameters and on effector concentrations. In fact, these panels reveal that such states can only exist at higher effector concentrations above approximately $10^{-4}$ M, and that they further rely on there being sufficiently weak activation.

Our results for a range of possible stable states demonstrate that by probing the dynamics at various effector concentrations, we can clearly distinguish systems that utilize effectors in distinct ways. The tuning knob we present in this study can therefore be viewed as a foundation for building fitted-model descriptions of expression data, and for gaining insights into mechanisms that both regulate transcription factor activity and would otherwise not be observable.

\section{Discussion}

A large body of experimental and theoretical work has studied a number of commonly observed motifs including simple two-gene switches, oscillators, and feed-forward networks, among others~\cite{Gardner:2000,Ozbudak:2004,HuangEnver:2007,Elowitz2000,Elowitz2004,HastyCollins2002,mangan2003structure,mangan2003coherent,mangan2006incoherent,carroll2001dna,Phillips:2020}. But even with these significant advances, there remains much more to explore. In fact, for {\it E. coli}, perhaps the most well-studied simple model organism, we still do not understand how more than $\sim$ 60\% of its genes are regulated~\cite{Santos-Zavaleta2019}. We also do not fully understand the mechanism by which effector molecules control the transcription factors responsible for gene regulation. As the field progresses in obtaining and interpreting high-throughput data, we will likely uncover additional aspects of gene regulation that require more nuanced modeling. Our work expands upon significant and valuable existing studies to consider a range of other motifs, and draws inspiration from the prevalence of induction in the biophysical literature as an experimental tool for tuning expression.

At the core of our analysis is a baseline model of three genes in which each gene represses the other two. By harnessing inducers as tuning knobs controlling the concentration of active repressors, we observe how tuning through an MWC thermodynamic interpretation of allosteric induction limits or broadens the scope of dynamics, and how robust (or sensitive) bifurcations separating different dynamic regimes are to changes in cooperativity and relative gene interaction strengths. 

In the first half of the paper we observe that the simple three-gene toggle switch, in which only one of the three genes can dominate expression in a stable state, follows a trajectory of decreasing dynamic complexity as inducer concentration(s) rise. There are two notable takeaways from analysis of the three resulting dynamic regimes. First, we observe through bifurcation diagrams that the inducer concentration thresholds at which systems shift between tristable and bistable dynamics change little as the strength of cooperativity increases, and that the maximum level of gene expression holds steady as inducer concentrations rise. This facilitates the fitting of model parameters to experimental data, since altering the $K_I$ of our $p_{\rm act}(c)$ function becomes the only way to noticeably shift the bifurcation threshold. Meanwhile, when not at a low cooperativity level, the inducer concentration at which the systems shift between different forms of tristable dynamics depends most strongly on perturbation of the models' regulatory interaction structures away from symmetry. By determining the probabilities of stabilizing at each of the possible three stable points given different initial conditions at low inducer concentrations, one can determine how close a system is to symmetry (i.e., equally strong repression among all genes). If a system tuned by one inducer is instead bistable at all inducer concentrations, it follows that this arises from a strongly skewed set of repression strengths explicitly favoring one gene, rather than from allosteric regulation.

While the baseline model demonstrates the utility of allosteric regulation as a tool to identify the physical parameters driving expression in experimental data, the second half of our study uses self-activation to highlight how the mechanism by which allostery controls transcription factor activity can significantly alter dynamics.
Introducing activators that competitively bind to sites along the DNA does not affect the stable expression levels observed in the baseline model, but decreases the probability of transitioning from one stable state to another by placing stable states in deeper potential wells within the potential landscape. When self-activation is sufficiently strong, or when activators and repressors bind non-exclusively along the DNA, it becomes possible for the system to stabilize while expressing more than one gene. Understanding the parameter conditions, including the ranges of effector concentrations, that allow different stable states to emerge is a useful tool for determining how cells tune activity to coordinate expression of multiple genes.

We also find across several examples with competitive and non-exclusive repressor/activator binding that the biological interpretation of the effector's role in regulation matters, whether an effector's binding determines activity or whether it directly determines function as a repressor or an activator. Considering systems in which activators compete with repressors to bind along the DNA, changing the effector's role does not alter either the types of dynamic phases possible or the system's response to increasing effector concentrations. In the case where the effector determines protein function, however, the effector concentrations at which bifurcations occur shift to higher values, and it becomes easier to transition out of an $\bar{R}_1$-dominant state and more difficult to transition out of an $\bar{R}_2$ or $\bar{R}_3$-dominant state. In non-exclusive binding settings, we find that the two interpretations of effector activity reveal distinct differences in the types of phases observed across parameter space, as well as in the ranges of effector concentrations at which different types of stable points are observed.

Tuning effector concentration(s) thus allows us to distinguish among biologically distinct models for a single gene regulatory motif, even when the particular set of observed stable states is characteristic of multiple models. For instance, in both the competitive (with strong activation) and non-exclusive binding settings, it is possible for the system to stabilize at low effector concentrations to any of the three single-gene-dominant states, or to a state suppressing all expression. As the effector concentration increases, however, the candidate models have divergent responses and are thus distinguishable. If the system becomes tristable to either suppress all expression or allow expression of a gene not targeted by the effector, the system identifies as competitive repressor-activator binding. If expression becomes suppressed entirely, the system instead allows non-exclusive binding, with the binding site determining transcription factor function and the effector only determining protein activity. Finally, if the system transforms to a bistable switch between no expression and expression of the effector-targeted gene alone, this implies non-exclusive binding where effector presence directly determines transcription factor function.

The mathematical interpretation of the biological mechanisms for induction embedded in the models we study have a significant impact on the types of dynamics observed. Our analysis therefore complements existing work on the direct incorporation of effector-driven transcription factor activity into models of gene regulation while further motivating the rigorous definition of \emph{how} effectors function. In so doing, the work presents a theoretical approach through which it is possible to uncover such properties, along with the model parameters that allow them, in conjunction with experiments. As evidenced by the diverse outputs we obtain from subjecting our three-noded networks to a range of effector inputs, there is also considerable richness in the input-output responses of these systems. Our findings underscore the importance and the challenges of understanding how such outcomes are realized in even more complex regulatory architectures, and invite dialogue with ongoing experimental efforts to reveal the full scope of the allosterome.

\begin{acknowledgments}
We are grateful to the NIH for support through award numbers DP1OD000217 (Director’s Pioneer Award) and NIH MIRA 1R35 GM118043-01. This research is funded in part by the Gordon and Betty Moore Foundation GBMF12214 (doi.org/10.37807/GBMF12214) to RP and RJR as part of the “Listening to Molecules” grant which has as its central mission to better understand the underlying mechanisms of allosteric regulation. RP is deeply grateful to the CZI Theory Institute Without Walls.
\end{acknowledgments}

\appendix

\section{Biologically permissible ranges for MWC parameters in allosteric induction} \label{app:MWC}

To determine the biologically permissible ranges for the MWC parameters defined in Eqn.~\ref{eq:pact} ($K_I$, $K_A$, and $\Delta\epsilon$), we assume that the absence of inducer does not imply the complete absence of inactive repressor, given that there may be some natural fractional error $q$. We suppose, then, that in the absence of inducer,
\begin{equation}
    p_{\text{act}}(c = 0) \geq 1 - q,
\end{equation}
or alternatively,
\begin{equation}
    1 - p_{\text{act}}(c = 0) = 1 - \frac{1}{1+e^{-\beta\Delta\epsilon}} = \frac{1}{1+e^{\beta\Delta\epsilon}} \leq q. \label{eq:cond1}
\end{equation}
Rearranging Eqn.~\ref{eq:cond1}, we obtain a condition for the energy difference $\Delta\epsilon$:
\begin{equation}
    e^{-\beta\Delta\epsilon} \leq \frac{q}{1-q}.\label{eq:cond1result}
\end{equation}
It follows that for the probability of activity to saturate in the absence of inducer ($q = 0$), the energy of the inactive repressor state must far exceed that of the active state so that $\Delta\epsilon \rightarrow \infty$ and the repressor is always active.

Gene expression is generally ``leaky,'' meaning that there is typically a basal level of expression even in the presence of a large repressor concentration. We assume that this leakiness is also reflected by the fraction $q$ such that a fraction of repressors less than $q$ are active in the limit of infinite inducer concentration, with
\begin{align}
    p_{A}(c\rightarrow\infty) &\approx \frac{\Big(\frac{c}{K_A}\Big)^m}{\Big(\frac{c}{K_A}\Big)^m + e^{-\beta\Delta\epsilon}\Big(\frac{c}{K_I}\Big)^m} \nonumber\\
    &= \frac{(K_I)^m}{(K_I)^m+e^{-\beta\Delta\epsilon}(K_A)^m} \leq q. \label{eq:cond2}
\end{align}
Rearranging Eqn.~\ref{eq:cond2}, we obtain a second condition
\begin{equation}
    \Big(\frac{K_A}{K_I}\Big)^m \geq \frac{1-q}{q}e^{\beta\Delta\epsilon}.\label{eq:cond2result}
\end{equation}
Eqns.~\ref{eq:cond1result} and~\ref{eq:cond2result} establish bounds on the dissociation constants $K_{A,\:I}$ and the energy difference $\Delta\epsilon$ for a given ``error'' $q$ and number of binding sites $m$.

Suppose, for example, that less than $10\%$ ($q = 0.1$) of repressors are active in the presence of an infinite inducer concentration. MWC parameters are then bounded by
\begin{equation}
    e^{-\beta\Delta\epsilon} \leq \frac{1}{9} \label{eq:boundsq01_1}
\end{equation}
and
\begin{equation}
    \Big(\frac{K_A}{K_I}\Big)^m \geq 9e^{\beta\Delta\epsilon}. \label{eq:boundsq01_2}
\end{equation}

We use these bounds to determine a permissible range of $K_I$ given fixed $K_A = 150 \:\mu M$ in Fig.~\ref{fig:prob}(B).

\section{Analytic bifurcations for the two inducer model}\label{sec:2indanalytic}
In this appendix, we show analytically that the linear thresholds separating the $(c_1, c_2)$ parameter space into tristable, bistable, and monostable regions in Section~\ref{sec:2ind} are the same as those in Section~\ref{sec:inducer1}.

The key bifurcation with respect to $c_1$ is the loss of the $\bar{R}_1$-dominant stable state in transitioning from tristable to bistable dynamics. Assuming that $\bar{R}_{2,3}^n\rightarrow 0$ as in the single inducer setting, Eqns.~\ref{eq:inducer2_r1} -~\ref{eq:inducer2_r3} indicate a steady state solution
\begin{align}
    \bar{R}_{ss} &= (\bar{R}_1, \bar{R}_2, \bar{R}_3)\nonumber\\
    &= \Big(\bar{a}, \frac{\bar{a}}{1+[p_{\text{act}}(c_1)\bar{a}]^n}, \frac{\bar{a}}{1+[p_{\text{act}}(c_1)\bar{a}]^n}\Big).
\end{align}
Since our assumptions require $\bar{R}_2 = \bar{R}_3 \leq \varepsilon$,
\begin{equation}
    \varepsilon \geq \frac{\bar{a}}{1+[p_{\text{act}}(c_1)\bar{a}]^n}
\end{equation}
and thus the $\bar{R}_1$ dominant steady state exists for
\begin{equation}
    p_{\text{act}}(c_1) \geq \frac{1}{\bar{a}}\sqrt[n]{\frac{\bar{a}}{\varepsilon} - 1},
\end{equation}
\\
\noindent the same as derived for a single inducer in Eqn.~\ref{eq:R1thresh}.

We derive the same bound for $p_{\text{act}}(c_2)$ by determining the bifurcation point with respect to $c_2$ at which the system no longer stabilizes to an $\bar{R}_2$-dominant state.

\section{Non-exclusive repressor-activator binding in the three-gene toggle switch}\label{app:nonexc}

\begin{figure*}
    \centering
    \includegraphics[width=0.9\textwidth]{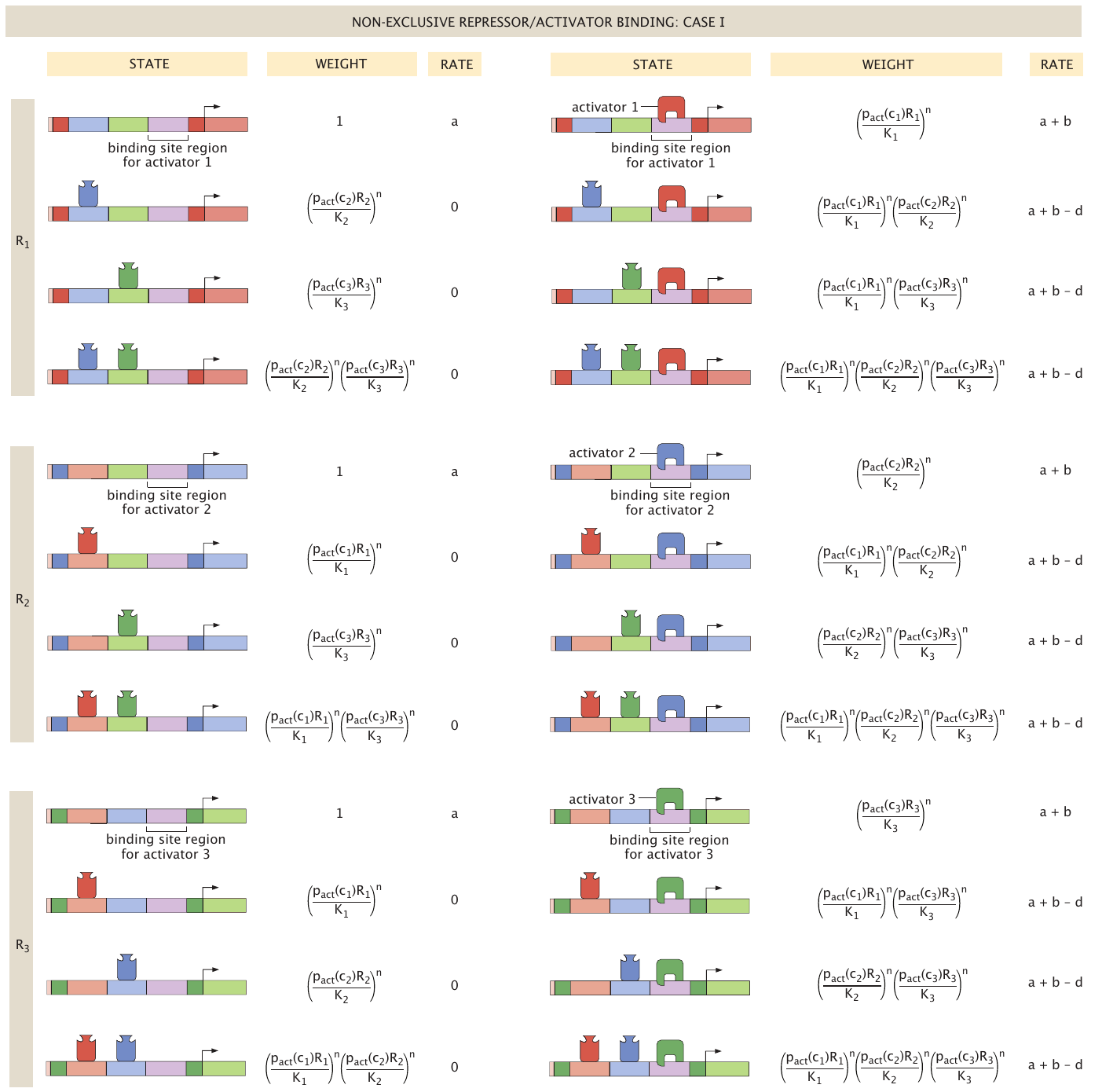}
    \caption{Expression of each protein $R_i$ in the three-gene toggle switch with self-activation and non-exclusive binding for repressors and activators. Effector binding is defined as in Fig.~\ref{fig:3geneselfact_effector}(A).}
    \label{fig:statesweights_nonexcA}
\end{figure*}

\begin{figure*}
    \centering
    \includegraphics[width=0.9\textwidth]{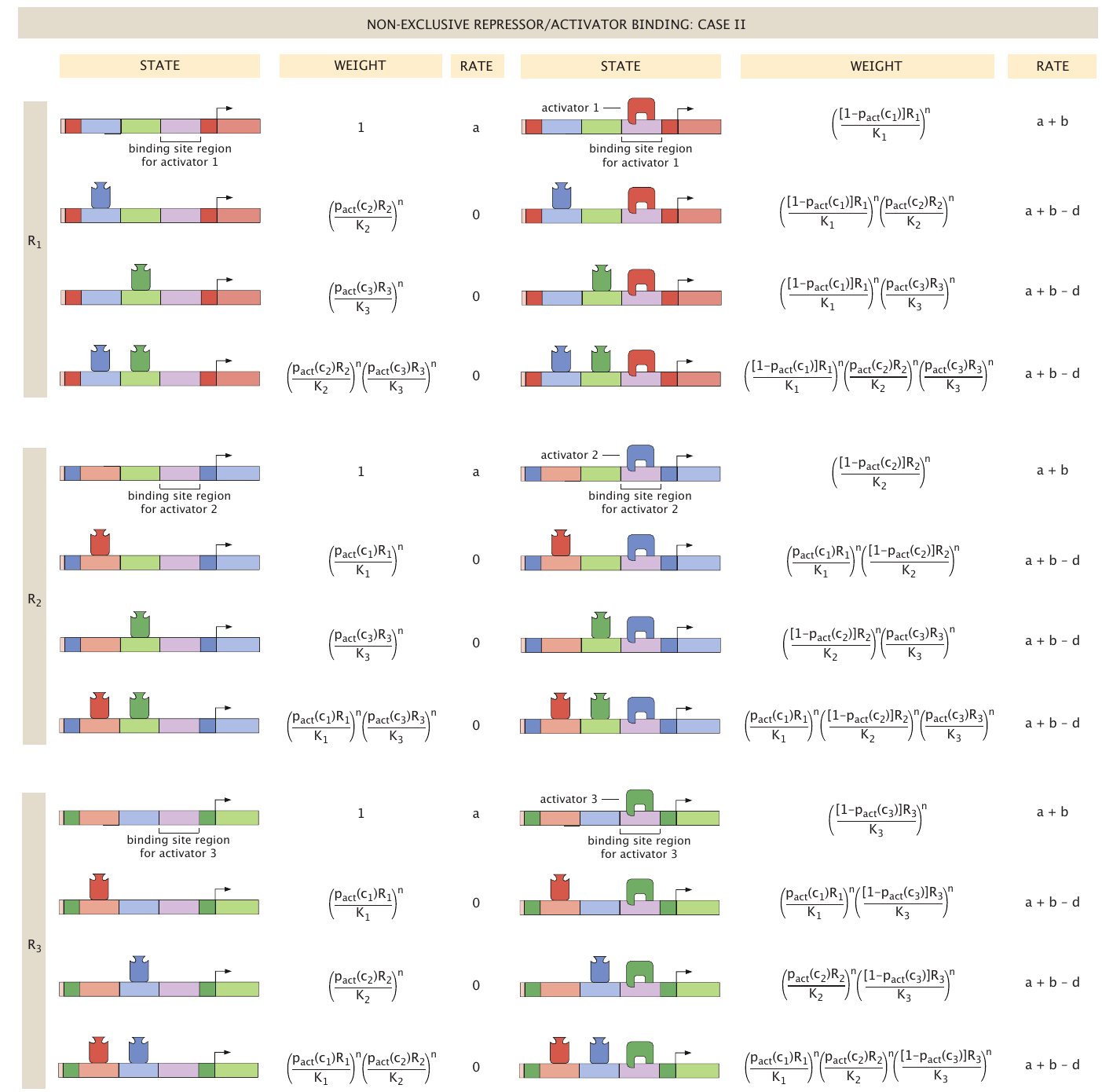}
    \caption{Expression of each protein $R_i$ in the three-gene toggle switch with self-activation and non-exclusive binding for repressors and activators. Effector binding is defined as in Fig.~\ref{fig:3geneselfact_effector}(B).}
    \label{fig:statesweights_nonexcB}
\end{figure*}

In Section~\ref{sec:nonexc} we discuss the three-gene toggle switch with self-activation where repressors and activators can bind non-exclusively at a gene's regulatory region. This appendix outlines the analytic model and how it is modified under different interpretations of effector activity.

Fig.~\ref{fig:statesweights_nonexcA} shows the thermodynamic states, weights and rates related to non-exclusive binding. Note that these include the same states describing competitive binding in Fig.~\ref{fig:statesweights_comp}(A), but with three additional states in which an activator and either one or two repressors can be bound simultaneously at the relevant gene regulatory site. In these additional states, the presence of repressor(s) dampens expression to $a + b - d$. The states and weights translate into a corresponding system of equations for the evolution of expression. Omitting the role of effector, the change in $R_1$ expression, for instance, is
\begin{widetext}
\begin{align}
    \frac{dR_{1}}{dt} &= \frac{a + (a+b)(\frac{R_{1}}{K_{1}})^n + (a+b-d)(\frac{R_{1}}{K_{1}})^n[(\frac{R_{2}}{K_{2}})^n + (\frac{R_{3}}{K_{3}})^n + (\frac{R_{2}}{K_{2}})^n(\frac{R_{3}}{K_{3}})^n]}{[1+(\frac{R_{1}}{K_{1}})^n][1+(\frac{R_{2}}{K_{2}})^n][1+(\frac{R_{3}}{K_{3}})^n]} - \frac{R_1}{\tau}\notag\\
    &= \frac{a}{[1+(\frac{R_{2}}{K_{2}})^n][1+(\frac{R_{3}}{K_{3}})^n]} + \frac{b(\frac{R_{1}}{K_{1}})^n}{1+(\frac{R_{1}}{K_{1}})^n} + \frac{(a-d)(\frac{R_{1}}{K_{1}})^n[(\frac{R_{2}}{K_{2}})^n+(\frac{R_{3}}{K_{3}})^n+(\frac{R_{2}}{K_{2}})^n(\frac{R_{3}}{K_{3}})^n]}{[1+(\frac{R_{1}}{K_{1}})^n][1+(\frac{R_{2}}{K_{2}})^n][1+(\frac{R_{3}}{K_{3}})^n]}-\frac{R_1}{\tau},\label{eq:nonexc_R1init}
\end{align}
where $a$ is the rate of protein production when gene transcription is unregulated, $a+b$ is the rate with bound self-activator, and $a+b-d$ is the previous rate attenuated by the presence of bound repressor. Note that the second line of Eqn.~\ref{eq:nonexc_R1init} rearranges the first to separate protein production into three physically meaningful terms: (i) protein expression in the presence of repressors only, (ii) expression exclusively due to activation, and (iii) expression in the presence of both repressors and activators. Rescaling variables in Eq.~\ref{eq:nonexc_R1init} and in the equivalent equations for $R_2$ and $R_3$ expression such that $\bar{R}_{i} = R_{i}/K_{1}$, $\bar{t} = t/\tau$, $\bar{a} = \tau a/K_{1}$, $\bar{b} = \tau b/K_{i}$, and $\bar{d} = \tau d/K_{1}$, we define our non-exclusive binding model in dimensionless form by the set of differential equations
\begin{align}
    \frac{d\bar{R}_{1}}{d\bar{t}} &= \frac{\bar{a}}{[1+(\frac{\bar{R}_{2}}{K^{(2)}})^n][1+(\frac{\bar{R}_{3}}{K^{(3)}})^n]} + \frac{\bar{b}\bar{R}_{1}^n}{1+\bar{R}_{1}^n} + \frac{(\bar{a}-\bar{d})\bar{R}_{1}^n[(\frac{\bar{R}_{2}}{K^{(2)}})^n+(\frac{\bar{R}_{3}}{K^{(3)}})^n+(\frac{\bar{R}_{2}}{K^{(2)}})^n(\frac{\bar{R}_{3}}{K^{(3)}})^n]}{[1+\bar{R}_{1}^n][1+(\frac{\bar{R}_{2}}{K^{(2)}})^n][1+(\frac{\bar{R}_{3}}{K^{(3)}})^n]} - \bar{R}_{1},\label{eq:barR1nonexc}\\
    \frac{d\bar{R}_{2}}{d\bar{t}} &= \frac{\bar{a}}{[1+\bar{R}_{1}^n][1+(\frac{\bar{R}_{3}}{K^{(3)}})^n]} + \frac{\bar{b}(\frac{\bar{R}_{2}}{K^{(2)}})^n}{1+(\frac{\bar{R}_{2}}{K^{(2)}})^n} + \frac{(\bar{a}-\bar{d})(\frac{\bar{R}_{2}}{K^{(2)}})^n\Big(\bar{R}_{1}^n+(\frac{\bar{R}_{3}}{K^{(3)}})^n+\bar{R}_{1}^n(\frac{\bar{R}_{3}}{K^{(3)}})^n\Big)}{[1+\bar{R}_{1}^n][1+(\frac{\bar{R}_{2}}{K^{(2)}})^n][1+(\frac{\bar{R}_{3}}{K^{(3)}})^n]} - \bar{R}_{2},\label{eq:barR2nonexc}\\
    \frac{d\bar{R}_{3}}{d\bar{t}} &= \frac{\bar{a}}{[1+\bar{R}_{1}^n][1+(\frac{\bar{R}_{2}}{K^{(2)}})^n]} + \frac{\bar{b}(\frac{\bar{R}_{3}}{K^{(3)}})^n}{1+(\frac{\bar{R}_{3}}{K^{(3)}})^n} + \frac{(\bar{a}-\bar{d})(\frac{\bar{R}_{3}}{K^{(3)}})^n\Big(\bar{R}_{1}^n+(\frac{\bar{R}_{2}}{K^{(2)}})^n+\bar{R}_{1}^n(\frac{\bar{R}_{2}}{K^{(2)}})^n\Big)}{[1+\bar{R}_{1}^n][1+(\frac{\bar{R}_{2}}{K^{(2)}})^n][1+(\frac{\bar{R}_{3}}{K^{(3)}})^n]} - \bar{R}_3,\label{eq:barR3nonexc}
\end{align}
with $\bar{d}\leq \bar{a} + \bar{b}$, $K^{(2)} = K_2/K_1$, and $K^{(3)} = K_3/K_1$. As in Section~\ref{sec:comp}, our analysis in Section~\ref{sec:nonexc} assumes equal binding affinities such that $K^{(2)} = K^{(3)} = 1$, and a maximal production rate $\bar{a} + \bar{b} = 2$.

Unlike the competitive binding case, the model defined in Eqns.~\ref{eq:barR1nonexc} -~\ref{eq:barR3nonexc} depends on \emph{two} independent parameters: (i) unregulated (basal) protein expression $\bar{a}$, with corresponding $\bar{b} = 2-\bar{a}$ denoting additional expression from activations, and (ii) the strength of repression from $\bar{d}$. We now explicitly incorporate the role of effector molecules in two ways.

First, Fig.~\ref{fig:statesweights_nonexcA} defines effector activity as described by Case I of Fig.~\ref{fig:3geneselfact_effector}(A), where the active concentration of a protein indexed by $i$ is $p_{\rm act}(c)\bar{R}_i$ regardless of its function as a repressor or an activator. For the system containing one effector at concentration $c$ that targets $\bar{R}_1$, the dimensionless Eqns.~\ref{eq:barR1nonexc} -~\ref{eq:barR3nonexc} in induction Case I become
\begin{align}
    \frac{d\bar{R}_{1}}{d\bar{t}} &= \frac{\bar{a}}{[1+\bar{R}_2^n][1+\bar{R}_3^n]} + \frac{\bar{b}[p_{\rm act}(c)\bar{R}_{1}]^n}{1+[p_{\rm act}(c)\bar{R}_{1}]^n} + \frac{(\bar{a}-\bar{d})[p_{\rm act}(c)\bar{R}_{1}]^n[\bar{R}_2^n+\bar{R}_3^n+\bar{R}_2^n\bar{R}_3^n]}{[1+[p_{\rm act}(c)\bar{R}_{1}]^n][1+\bar{R}_2^n][1+\bar{R}_3^n]} - \bar{R}_{1},\label{eq:barR1nonexcA}\\
    \frac{d\bar{R}_{2}}{d\bar{t}} &= \frac{\bar{a}}{[1+[p_{\rm act}(c)\bar{R}_{1}]^n][1+\bar{R}_3^n]} + \frac{\bar{b}\bar{R}_2^n}{1+\bar{R}_2^n} + \frac{(\bar{a}-\bar{d})\bar{R}_2^n\Big([p_{\rm act}(c)\bar{R}_{1}]^n+\bar{R}_3^n+[p_{\rm act}(c)\bar{R}_{1}]^n\bar{R}_3^n\Big)}{[1+[p_{\rm act}(c)\bar{R}_{1}]^n][1+\bar{R}_2^n][1+\bar{R}_3^n]} - \bar{R}_{2},\label{eq:barR2nonexcA}
\end{align}
\begin{align}
    \frac{d\bar{R}_{3}}{d\bar{t}} &= \frac{\bar{a}}{[1+[p_{\rm act}(c)\bar{R}_{1}]^n][1+\bar{R}_2^n]} + \frac{\bar{b}\bar{R}_3^n}{1+\bar{R}_3^n} + \frac{(\bar{a}-\bar{d})\bar{R}_3^n\Big([p_{\rm act}(c)\bar{R}_{1}]^n+\bar{R}_2^n+[p_{\rm act}(c)\bar{R}_{1}]^n\bar{R}_2^n\Big)}{[1+[p_{\rm act}(c)\bar{R}_{1}]^n][1+\bar{R}_2^n][1+\bar{R}_3^n]} - \bar{R}_3.\label{eq:barR3nonexcA}
\end{align}
Evaluating the fixed points of Eqns.~\ref{eq:barR1nonexcA} -~\ref{eq:barR3nonexcA}, we obtain the results discussed in Section~\ref{sec:nonexcA}.
\vspace{0.1cm}

An alternative interpretation for effector activity, described by Case II of Fig.~\ref{fig:3geneselfact_effector}(B), transforms the available states and weights to those in Fig.~\ref{fig:statesweights_nonexcB}, where the concentration of $\bar{R}_i$ is scaled by $p_{\rm act}(c)$ when repressing expression of other proteins, and scaled by $1 - p_{\rm act}(c)$ when activating its own expression. Eqns.~\ref{eq:barR1nonexc} -~\ref{eq:barR3nonexc} then become

\begin{align}
    \frac{d\bar{R}_{1}}{d\bar{t}} &= \frac{\bar{a}}{[1+\bar{R}_2^n][1+\bar{R}_3^n]} + \frac{\bar{b}[(1-p_{\rm act}(c))\bar{R}_{1}]^n}{1+[(1-p_{\rm act}(c))\bar{R}_{1}]^n} + \frac{(\bar{a}-\bar{d})[(1-p_{\rm act}(c))\bar{R}_{1}]^n[\bar{R}_2^n+\bar{R}_3^n+\bar{R}_2^n\bar{R}_3^n]}{[1+[(1-p_{\rm act}(c))\bar{R}_{1}]^n][1+\bar{R}_2^n][1+\bar{R}_3^n]} - \bar{R}_{1},\label{eq:barR1nonexcB}\\
    \frac{d\bar{R}_{2}}{d\bar{t}} &= \frac{\bar{a}}{[1+[p_{\rm act}(c)\bar{R}_{1}]^n][1+\bar{R}_3^n]} + \frac{\bar{b}\bar{R}_2^n}{1+\bar{R}_2^n} + \frac{(\bar{a}-\bar{d})\bar{R}_2^n\Big([p_{\rm act}(c)\bar{R}_{1}]^n+\bar{R}_3^n+[p_{\rm act}(c)\bar{R}_{1}]^n\bar{R}_3^n\Big)}{[1+[p_{\rm act}(c)\bar{R}_{1}]^n][1+\bar{R}_2^n][1+\bar{R}_3^n]} - \bar{R}_{2},\label{eq:barR2nonexcB}\\
    \frac{d\bar{R}_{3}}{d\bar{t}} &= \frac{\bar{a}}{[1+[p_{\rm act}(c)\bar{R}_{1}]^n][1+\bar{R}_2^n]} + \frac{\bar{b}\bar{R}_3^n}{1+\bar{R}_3^n} + \frac{(\bar{a}-\bar{d})\bar{R}_3^n\Big([p_{\rm act}(c)\bar{R}_{1}]^n+\bar{R}_2^n+[p_{\rm act}(c)\bar{R}_{1}]^n\bar{R}_2^n\Big)}{[1+[p_{\rm act}(c)\bar{R}_{1}]^n][1+\bar{R}_2^n][1+\bar{R}_3^n]} - \bar{R}_3,\label{eq:barR3nonexcB}
\end{align}
\end{widetext}
\noindent describing a system containing only one effector at concentration $c$ targeting $\bar{R}_1$ activity. Evaluating the fixed points of Eqns.~\ref{eq:barR1nonexcB} -~\ref{eq:barR3nonexcB}, we obtain the results discussed in Section~\ref{sec:nonexcB}.

\section{Code availability}\label{app:code}
All Matlab code used to generate graphs in figures throughout this work are available \footnote{Supporting Matlab code is available on GitHub at \href{https://github.com/Rebecca-J-Rousseau/inducible3genetoggle}{github.com/Rebecca-J-Rousseau/inducible3genetoggle}}.

\bibliography{main}

\end{document}